\documentclass[aps,prl,twocolumn,groupedaddress,fleqn,showkeys]{revtex4-1}
\usepackage{graphicx}
\usepackage{epstopdf}
\usepackage[fleqn]{amsmath}
\usepackage{txfonts}
\usepackage{eqnarray}
\usepackage{bm}
\usepackage{xr}
\usepackage{xcolor}
\usepackage{eucal}
\usepackage{amssymb}
\usepackage{mathrsfs}
\usepackage{multirow}

\begin{document}

\title{Influence of elastic strain on the thermodynamics and kinetics of lithium vacancy in bulk LiCoO$_2$}

\author{Ashkan Moradabadi}
\email[ashkan.moradabadi@fu-berlin.de]{}
\affiliation{Institut f\"ur Chemie und Biochemie, 
Freie Universit\"at Berlin, Takustr. 3, 14195 Berlin, Germany} 
\affiliation{Institut f\"ur Materialwissenschaft, Fachgebiet Materialmodellierung, 
Technische Universit€at Darmstadt, Jovanka-Bontschits-Str. 2, 64287 Darmstadt,
Germany}
\author{Payam Kaghazchi}
\affiliation{Institut f\"ur Chemie und Biochemie, 
Freie Universit\"at Berlin, Takustr. 3, 14195 Berlin, Germany}
\author{Jochen Rohrer}
\affiliation{Institut f\"ur Materialwissenschaft, Fachgebiet Materialmodellierung, 
Technische Universit€at Darmstadt, Jovanka-Bontschits-Str. 2, 64287 Darmstadt,
Germany}
\author{Karsten Albe}
\email[albe@mm.tu-darmstadt.de]{}
\affiliation{Institut f\"ur Materialwissenschaft, Fachgebiet Materialmodellierung, 
Technische Universit€at Darmstadt, Jovanka-Bontschits-Str. 2, 64287 Darmstadt,
Germany}
\date{\today}
\vspace{0.5cm}


\begin{abstract}
\section{Abstract}
The influence of elastic strain on the lithium vacancy formation and migration in bulk LiCoO$_2$
is evaluated by means of first-principles calculations within density functional theory (DFT). Strain dependent energies are determined directly from  defective cells and also within linear elasticity theory from the elastic dipole tensor ($G_{ij}$) for ground state and saddle point configurations.
 We analyze finite size-effects in the calculation of $G_{ij}$, compare the predictions of the linear elastic model with those obtained from direct calculations of defective cells under strain and discuss the differences. 
Based on our data,  we calculate the variations in vacancy concentration and mobility due to the presence of external strain in bulk $\rm LiCoO_2$ cathodes.
Our results reveal that elastic in-plane and out-of-plane strains can 
significantly change the ionic conductivity
of bulk LiCoO$_2$ by an order of magnitude and thus strongly affect the performance of Li-secondary batteries. 
\end{abstract}


\maketitle
\clearpage


\section{I. Introduction}
The presence of strain fields can significantly influence the efficiency and lifetime of functional materials such as semiconductors, solar cells and Li-ion batteries \cite{Sun,Penev,Hubbard,yi,cui,rutooj,seif,malik,wang,karsten3}.
In principle, strain can be internally generated by structural defects  within the bulk and/or interfaces or can be induced by external loads.
These strains or corresponding stresses can cause lattice deformations and distortions and therefore also affect the formation and migration of point defects. As a result, the conductivity of an ionic conductor can be significantly changed \cite{Walu,Azp,Fisher,Freedman,Goyal}.
This coupling is most relevant in Li-ion batteries, where due to charging/discharging processes, bulk and  interfacial as well as thermal strains can occur. 
The fact that induced stresses, which raise during intercalation, could weaken the interface between electrode/electrolyte and finally degrade the battery performance has been extensively discussed in the past \cite{ashkan,ashkan2,khang1,khang2,khang3,amine,sung,ning,Bower,garcia,Fisher2,Okubo}. Much less is known, however, about the coupling of strain fields to the formation and migration of Li vacancies in the bulk part of the material.

As one of the first, Choi $et$ $al.$ \cite{choi} have examined the effect of intercalation-induced stress on Li migration in LiCoO$_2$ 
using experimental techniques such as electrochemical impedance spectroscopy together with cyclic voltammetry. They showed that the mismatch strain between intercalated and deintercalated states in 
LiCoO$_2$ is due to a phase transition between the $\alpha$ and $\beta$ phases. This results in intercalation-induced stress in the phase boundary region. 
Since an inserted Li ion causes structural disorder in the host material, it introduces a strain field which affects the next intercalated Li and leads to a 
strain-induced elastic interaction between all intercalated Li ions. This elastic interaction has both short and long range effects and it was
 also found that as the particle size decreases, the stress gradient across LiCoO$_2$ particles increases.

Transmission electron microscopy (TEM), X-ray diffraction and electrochemical strain microscopy (ESM) have revealed that upon delithiation the c-axis  in Li$_x$CoO$_2$ (x=0.5) expands by about 2\% \cite{wang2,Takahashi} and rather large stresses occur during charging/discharging process at different charge/discharge rates (c-rates) \cite{Diercks,yong,sheldon,Balke}.
Garcia $et.$ $al.$, for example, have shown based on a continuum model that large stress values (up to $\pm$ 200 MPa) are generated at particle contacts and these values increase with increasing  c-rate of discharge \cite{garcia}.
Xiong $et$ $al.$  have shown in an {\em ab-initio} based study that during deintercalation of Li$_x$CoO$_2$, the $c$-axis increases up to 3.25\% \cite{ouyang2}. 
Li $et~al.$ also reported that compressive stress raises while the Li concentration 
is decreasing and at x=0.5, the measured strain at the O-Co-O octahedral slabs is 4.8\% \cite{Li}.
A mathematical model was developed by Renganathan $et$ $al.$ to reveal the mechanical stresses generated during the discharge process in carbon and LiCoO$_2$ \cite{Renganathan}. Their findings show that 
at high discharge c-rates, stresses also increase and the stress caused by phase transformations is related to the amount of each phase present in the electrode. They also  concluded that particle 
size and distribution can affect the generated stress~\cite{Renganathan}.
Critical rates of charging and particle size below which fracture of LiCoO$_2$ occurs were predicted by Zhao $et$ $al.$ using a kinetic and fracture mechanics model. They 
have shown that as the discharge rate increases the particle size should decrease so that fracture is prevented (for c-rates more than 5 C, particle sizes less than 200 nm) \cite{zhao}.  

While all these studies point to the fact that significant stress and strain levels can occur in cathode materials of Li-secondary batteries, their influence on Li-ion diffusion has hardly been studied.
In a recent theoretical work Ning $et$ $al.$ \cite{ning} 
 have shown that  by applying uniaxial tensile strain along the c-axis of bulk LiCoO$_2$, the Li diffusion barrier decreases~\cite{ning}, but did not derive consider the case of a more complex tensorial strain field.

In this study, we calculate the elastic dipole tensor~\cite{Leibfried} in order to characterize the coupling of  strain fields to the formation and migration energies of Li vacancy in bulk-LiCoO$_2$. 
We obtain the components of the elastic dipole tensor for defect formation and also migration from total energy calculations within density functional theory. 
Similar calculations of the defect dipole tensor have been recently reported for defects in metals and silicon by Varvenne $et \, al.$ \cite{Varvenne2013} and in UO$_2$ by Goyal $et \, al.$ \cite{Goyal}. 
In order to demonstrate the degree of coupling between lateral and longitudinal components of the stress tensor in LiCoO$_2$, various supercell geometries are studied. Moreover, we compare the predictions from linear elasticity theory with directly calculated formation and migration energies of vacancy under strain and show that the elastic dipole-tensor allows us to quantify the influence of strain in a computationally efficient manner. Finally, we estimate the effect of lateral and longitudinal strains on ionic conductivity in bulk LiCoO$_2$.

\section{II. Theory and Computational methods}
\subsection{Elastic dipole tensor}

The insertion of a point defect into a crystal produces local elastic distortions.
Moreover, there will be an interaction between this defect and a stress or strain field present in the crystal. This is similar to the interaction of an electric dipole with an applied electric field. Therefore, a defect inducing local distortions is called an elastic dipole, which is -contrary to the electric dipole- characterized by a second-rank tensor. This elastic dipole tensor, which is also called double force tensor \cite{Leibfried}, is the negative derivative of the defect formation energy $E_{d}$ with respect to an imposed bulk strain \\
$
G_{ij}=-\dfrac{\partial E_d}{\partial \epsilon_{ij}},
$\\
if, as usual, we ignore  entropy contributions.  Thus, $G_{ij}$  is relating the atomic structure of a point defect and its elastic field.
In case of a purely dilatational strain the defect relaxation volume  $\Delta V= \dfrac{1}{3B} {\rm Tr} \lbrace G_{ij} \rbrace$, where $B$ is the bulk modulus, can be directly obtained from $G_{ij}$. In principle,
the concept of the elastic dipole tensor can be conveniently understood by expanding the free energy per volume 
in terms of the density of defects $n_d=N_d/V$ and the strain tensor $\epsilon_{ij}$ (both being intensive quantities) \cite{gillan3}. If entropy contributions are neglected the expansion of the  energy density (T=0 K) reads as: 

\begin{flalign}
E(n_d, \epsilon) &= E_{o}+ 
                  \underbrace{\sum_{i,j} 
                   \frac{\partial E}{\partial \epsilon_{ij}}}_{\sigma_{ij}=0} \epsilon_{ij}
                   +\frac{1}{2}\sum_{i,j}\frac{\partial^2 E}{\partial \epsilon_{ij}\partial\epsilon_{kl}}\epsilon_{ij}\epsilon_{kl}   \nonumber \\ & + \frac{\partial E}{\partial n_d} n_d + \sum_{i,j}\frac{\partial^2 E}{\partial n_d \partial \epsilon_{ij}}\epsilon_{ij}n_d+ \ldots \nonumber \\                      
                    &=E_{o}  + 
                                       \frac{1}{2}\sum_{i,j}C_{ijkl}\epsilon_{ij}\epsilon_{kl}
                                        + n_d \left ( E_d + \sum_{i,j}\frac{\partial^2 E}{\partial \epsilon_{ij}\partial n_d }\epsilon_{ij}\right)+ \ldots. \nonumber\\
                   &=E_{o} +                                   \frac{1}{2}\sum_{i,j}C_{ijkl}\epsilon_{ij}\epsilon_{kl}
                     + n_d \left ( E_d + \sum_{i,j}\frac{\partial \sigma_{ij}}{ \partial n_d }\epsilon_{ij}\right)+ \ldots. \nonumber\\
                     &=E_{o} +                                   \frac{1}{2}\sum_{i,j}C_{ijkl}\epsilon_{ij}\epsilon_{kl}
                                          + n_d \left ( E_d - \sum_{i,j}{ G_{ij}}\epsilon_{ij}\right)+ \ldots.&&
\label{eq:energy_1}
\end{flalign}

Here $E_o$ is the total energy of the non-strained defect-free system,  $n_d$ the  number of defects, respectively, $E_d$ is the formation energy of a defect,  $\sigma_{ij}$ is the  stress, $C_{ijkl}$ are the components of the stiffness tensor and $G_{ij}$
is the elastic dipole tensor. As such $G_{ij}$ describes the interaction of the defect with a strain field. The change in  energy under strain that is exclusively  due to the presence of a defect is given by

\begin{align} \label{eq1-1}
\Delta E=-\sum\limits_{ij} G_{ij}\epsilon_{ij}.
\end{align} 
Thus, for example, a positive lattice strain would lower the formation energy of a defect having a positive relaxation volume.
The stress in a material under strain $\epsilon_{ij}$ with a defect density $n_d$ is eventually given by

\begin{align}
  \sigma^d_{ij} \equiv \frac{\partial E(n_d,\epsilon_{ij})}{\partial \epsilon_{ij}} =  {\sum_{kl}C_{ijkl}\epsilon_{kl}}-n_dG_{ij}= \sigma^0_{ij}-n_d G_{ij}
  \label{eq4},
\end{align}
if we only consider first-order components of the elastic dipole tensor.
Note that -depending on the stress definition- different sign conventions have been proposed in literature. Here we stick to the original one used by Leibfried and Breuer \cite{Leibfried}.

With relation (\ref{eq4}) it is straightforward to compute the components of the elastic dipole tensor numerically using atomistic methods.   
In a given periodic supercell, an individual defect is introduced. While the cell parameters are fixed, the atomic positions are relaxed and the induced stress is calculated. Then, the elastic dipole tensor can be obtained from the relation

\begin{align}
G_{ij} = -\frac{\partial E_d}{\partial \epsilon_{ij}}= -\left. \frac{\partial \sigma_{ij}}{\partial n_d}\right |_{\epsilon_{ij}} =- \frac{1}{n_d} (\sigma_{ij}^{d} - \sigma_{ij}^0) = -V_0\Delta \sigma_{ij},
\label{eq:Gij}
\end{align}
where $V_0$ is the volume of the supercell containing one defect, $\sigma_{ij}^{d}$ is the stress of the defective cell
and $\sigma_{ij}^0$ is the stress of the defect-free cell (which in principle should be or very close to zero).

In supercell calculations, there will be a certain unwanted contribution to
the energy described in Eq. \ref{eq:energy_1} from the interaction between the defects in their periodic images. It is therefore necessary, either to correct for this interaction, or to increase the size of the repeating unit to make it negligible. 
In the case of charged defects, another contribution comes from the Coulomb interaction between the defects, which must be corrected  as shown by Leslie and Gillan \cite{gillan3}.
Elastic contributions can usually be made small enough by increasing the cell size. If {\it ab-initio} methods are used, however, the accessible cell sizes are rather limited and elastic interactions can induce higher order effects, that are not covered by linear elasticity. 
As can be seen from Eq.~\ref{eq:Gij}, the variation of stress by a defect follows the relation $\Delta\sigma_{ij} = -G_{ij}/V_o$ and thus goes to zero in the dilute limit
($\lim_{V_o\rightarrow\infty}\Delta\sigma_{ij}= 0$).
Therefore, by plotting the components of $\Delta\sigma_{ij}$ as function of the inverse volume, one can test the convergence behavior of the calculated stresses. 
For small supercell sizes higher order effects will affect the linearity of $\Delta\sigma_{ij}$ with respect to the inverse volume.
In practice, a polynomial fit including higher order terms are required. The coefficient of the linear term dominates at small values of inverse volume and thus still corresponds to the elastic dipole tensor.

\subsection{Defect formation and migration energies under strain}

The defect formation energy for a given homogeneous strain $\epsilon_{ij}$, considering Eq. \ref{eq1-1}, is then given by 
\begin{equation}
E_d (\epsilon_{ij})= E_d(0)-\sum_{ij} G_{ij} \epsilon_{ij}.
\label{eq:Ed_indirect}
\end{equation}
Note, that $E_d (\epsilon_{ij})$ can also be directly calculated as  shown later.

In the present work, we only consider a neutral Li vacancy. The reason is that we want to disentangle the  electrostatic  and elastic interactions.
The conventional way to calculate the formation energy of defects in the neutral state and in the presence of strain is to use the following equation \cite{Zhang1991}
\begin{eqnarray}
\label{eq:Ed_direct}
E_{d}(\epsilon_{ij})&=& {E}_{\rm tot}^{\rm (Li-vacancy)}(\epsilon_{ij})-E^{\rm p}_{\rm tot}(\epsilon_{ij})+\mu_{Li}.
\end{eqnarray}
In this equation, the first term is the total energy of the strained system containing a single neutral Li vacancy, the second term is the total energy of the strained pristine system and the last term is the strain-free chemical potential of the Li reservoir.  
This can be compared to the defect formation energy as a function of strain  calculated from Eq.~\ref{eq:Ed_indirect} for a specific concentration. 
It should be noted here that the vacancy formation energy depends on the chemical potential of lithium in the reservoir, which is taking up the removed Li atom. Thus, in principle one has to consider the fact that strain might also affect the reservoir.


We now move to the  coupling of defect migration and strain fields. The energy barrier that is required for an ion to jump between two sites is obtained by the energy difference between the saddle point and the initial configuration

\begin{equation}\label{eq12}
 E_{b}(\epsilon)=E^{S}(\epsilon_{ij})-E_d(\epsilon_{ij}),
\end{equation}

where $E^{S}$ is the energy of the defective system in the saddle point (transition state) and $E_d$ the  energy of the defective initial state.

For a system under strain, the energy barrier can be computed by applying a particular strain to the simulation  cell and direct calculation of this energy difference. 
Alternatively, one can also calculate the defect dipole tensor $G^S_{ij}$ at the saddle point configuration and obtain the strain dependent barrier from

\begin{equation}\label{eq13}
  \begin{aligned}
     E_{b}(\epsilon_{ij})&= E_{b}(0)- \sum_{ij} \left (G_{ij}^{S}\epsilon_{ij}-G_{ij}\epsilon_{ij}\right)\\\
    &= E_{b}(0)- \sum_{ij}\Delta G_{ij}\epsilon_{ij},
  \end{aligned}
\end{equation}
where  $E_{b}(\epsilon_{ij})$ is the migration energy barrier in the strained material, 
$ E_{b}(0)$ is the migration energy barrier in the unstrained material, while
$\Delta$G$_{ij}^{b}$ is the change of the elastic dipole tensor by going from the initial to the transition state. For calculating the $\Delta$G$_{ij}$, 
we perform two single point calculations, one for the initial and one for the transition state in order to obtain $\Delta\sigma^d_{ij}$. Afterwards, using 
Eq. \ref{eq:Gij}, the elastic dipole tensor for the initial and transition state can be calculated. The atomic coordinations used in the single point calculations 
are obtained from the NEB calculation for the unstrained case. Therefore, by performing only one NEB calculation at $\epsilon_{ij}=0$, the strain-dependent
activation barriers for Li-diffusion can be investigated.


\begin{figure}[!th]
	\includegraphics[width=0.49\textwidth]{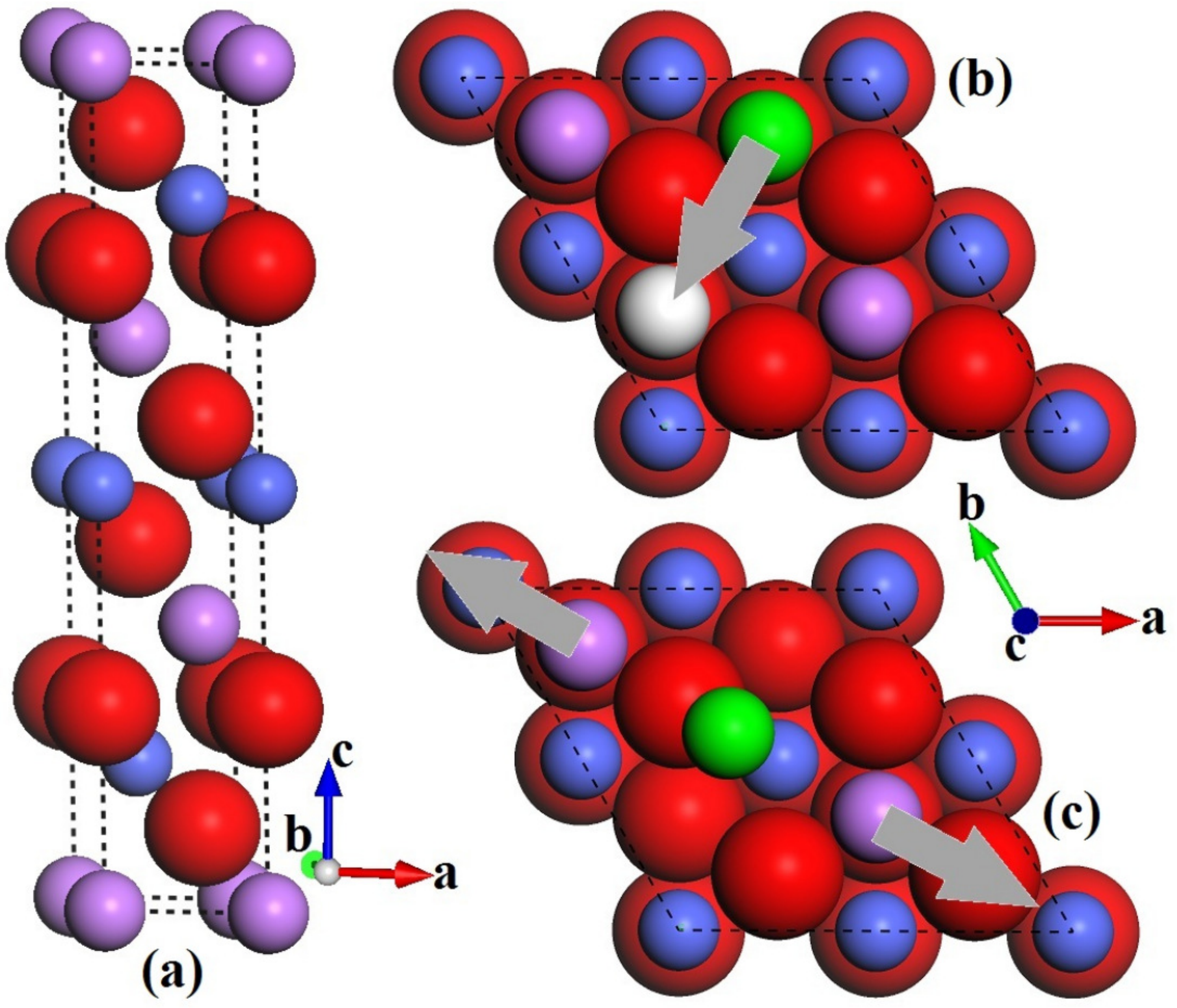}
	\caption{(color online) (a) 3D view of conventional 1$\times$1$\times$1 unit cell for bulk LiCoO$_2$, (b) top view of Li diffusion 
		through a single-vacancy mechanism on a direct pathway and (c) diffusing Li position in the saddle point while causing distortion for its nearest Li neighbors. 
		The direction of the gray arrows determined according to the sign of G (which is equal to the opposite sign of stress). Li, O and Co are shown with violet, red and blue, respectively, while migrating Li and Li vacancy are shown with green and white, respectively.}
	\label{str}
\end{figure}

\subsection{Model structures}

\begin{figure*}[!th]
	\includegraphics[width=0.99\textwidth]{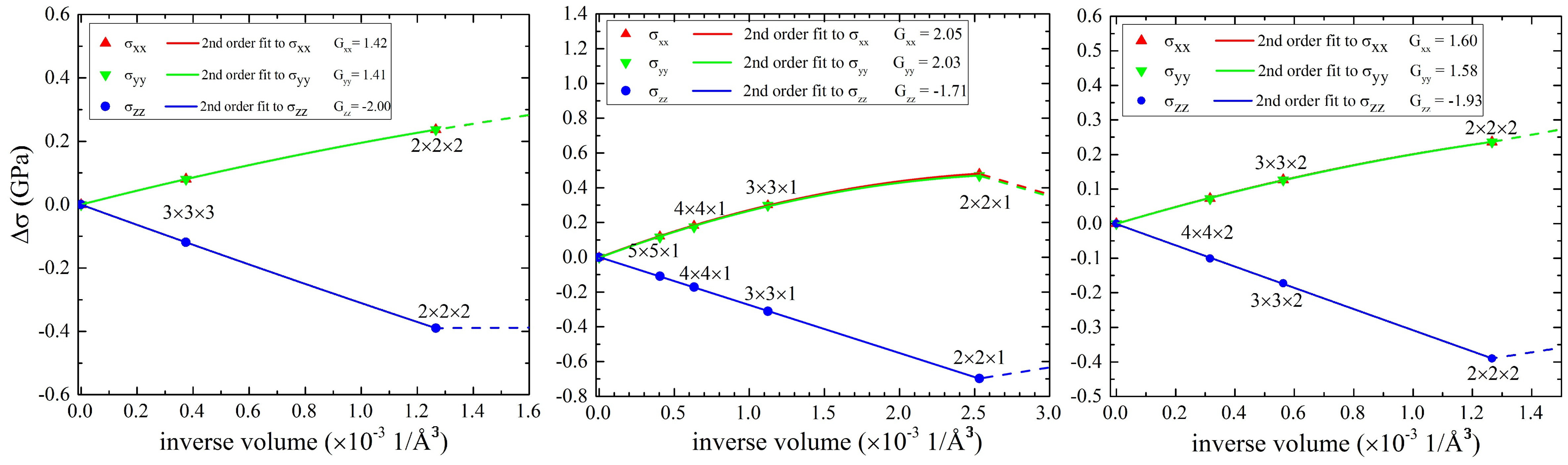}
	\caption{(color online) Variations of the diagonal elements of the stress tensor for various supercell sizes and aspect ratios.
		The stress variation is fitted to a second-order polynomial in the range indicated by solid lines; the coefficients of the linear term which corresponds to the elastic dipole tensor in the dilute limit are given in the legend. Supercell sizes for the left figure are $ n \times n \times n$ ($ 1 \leq n \leq 3 $), middle figure  $n \times n \times 1$ ($ 1 \leq n \leq 5 $) and right figure $n \times n \times 2$ ($ 1 \leq n \leq 4$). The $(0,0)$ point is included in all plots. }
	\label{lateral}
\end{figure*}

\begin{figure*}[!th]
	\includegraphics[width=0.99\textwidth]{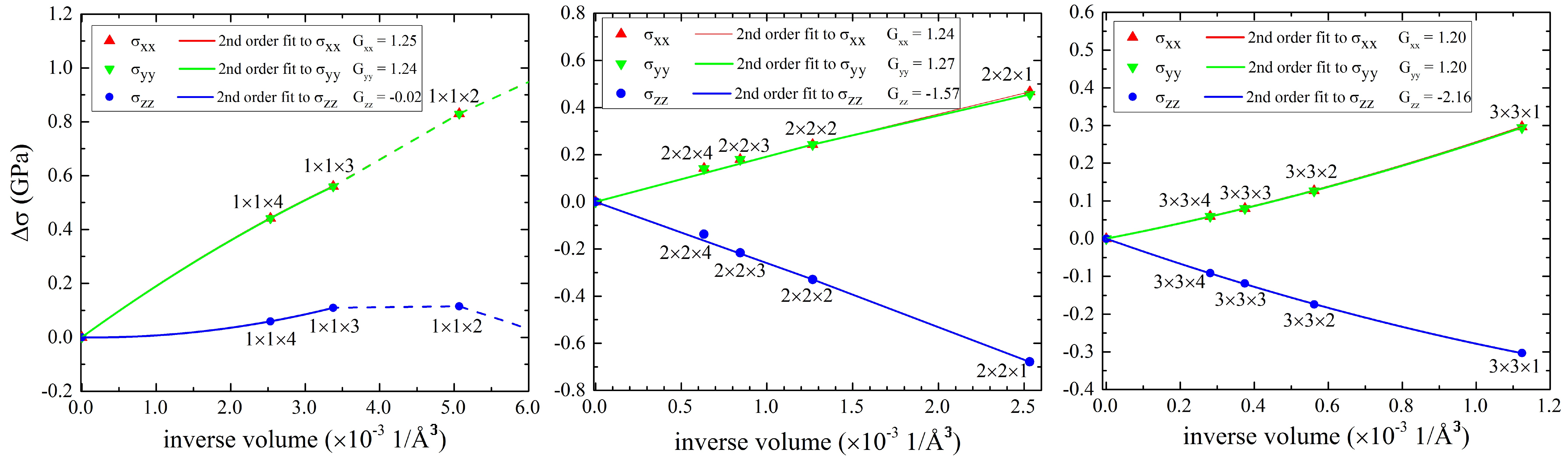}
	\caption{(color online)  
		Variations of the diagonal elements of the stress tensor.
		The stress variation is fitted to a second-order polynomial in the range indicated by solid lines;
		the coefficients of the linear term which corresponds to the elastic dipole tensor in the dilute limit are given in the legend.
		Supercell sizes for left figure are $1\times 1 \times n$, middle figure is $2 \times 2 \times n$ and right figure is $3 \times 3 \times n$ ($ 1\leq n \leq 4$). The $(0,0)$ point is included in all plots.}
	\label{longitudinal}
\end{figure*}

\begin{figure}
	\includegraphics[width=0.4\textwidth]{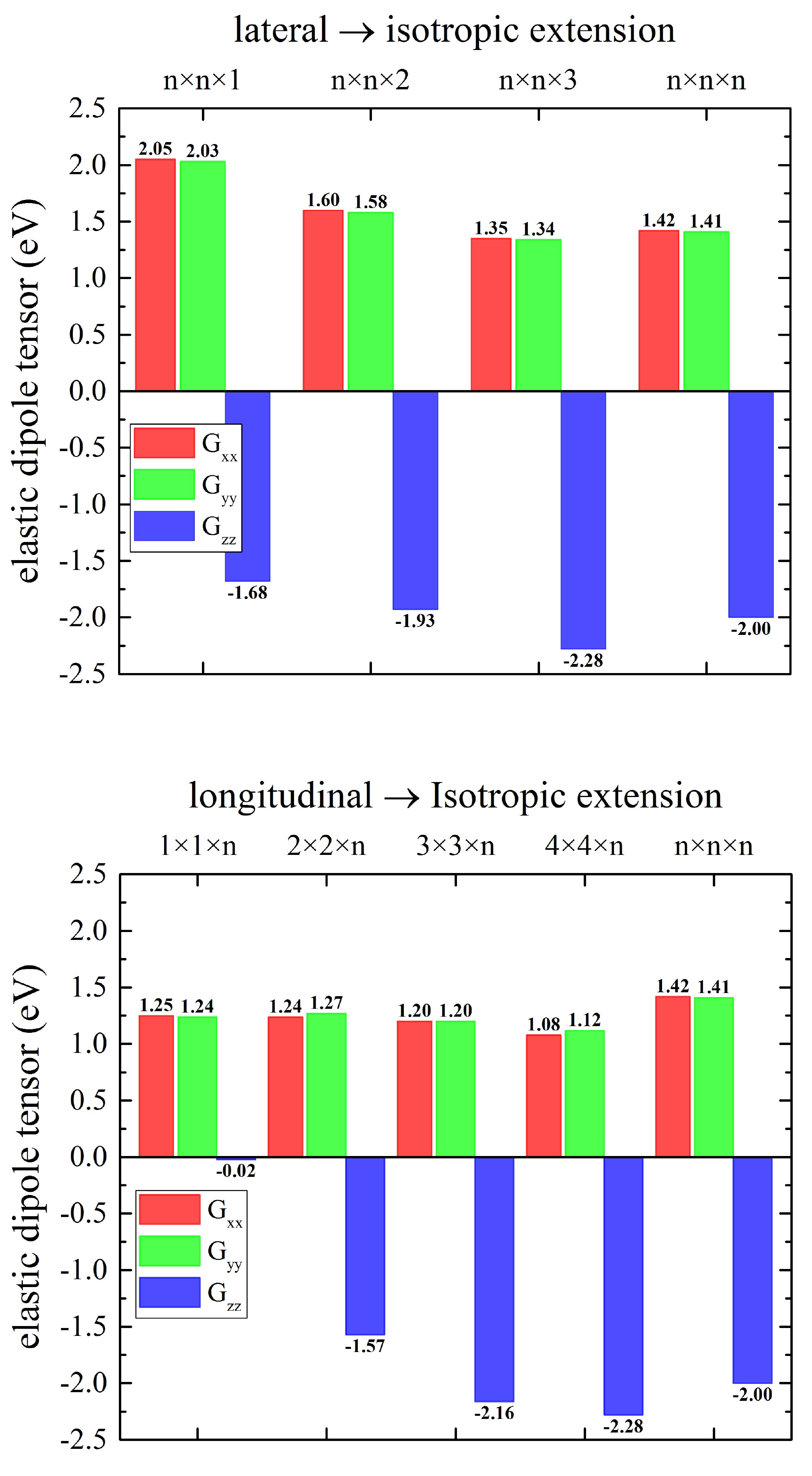}
	\caption{(color online) Histogram plot of the size effect analysis for  the diagonal components of the elastic dipole tensor in bulk LiCoO$_2$ containing a single Li vacancy.}
	\label{histo}
\end{figure}

Fig. \ref{str} (a) shows the  unit cell of bulk LiCoO$_2$. Li atoms (violet color) are coordinated within the octahedrals of CoO$^{-2}$. 
During the delithiation, Li vacancies are formed and start to migrate. Considering the layered structure of LiCoO$_2$, we 
expect a significant influence of strain on the formation and migration of these vacancies. The strain can either originate from the expansion/contraction during 
the deintercalation/intercalation, respectively, or can be applied externally (e.g. interfacial strains from SEI, solid electrolyte or binder).

In Fig. \ref{str}(b), the direct mechanism of Li diffusion towards the single vacancy is indicated (or migration of vacancy towards Li). During this process, the Li ion
must overcome an energy barrier $E_b$ by pushing the two nearby Li ions and passing through them. This mechanism is shown in Fig. \ref{str}(b) and (c) by gray arrows. Therefore, Li on the saddle point is under compressive stress (tensile on its neighbors).  
While there are also other Li migration mechanisms reported \cite{VanderVen,ashkan},   in the following, we  will focus on the strain dependence of the formation and migration of a neutral single Li vacancy moving on a direct pathway. Since we are only interested in the strain dependency, we deliberately study neutral cells in order to disentangle electrostatic image interactions of the charged defects from the elastic image interactions being also present. 

\subsection{Computational details}
The calculations were performed using the local atomic-orbital DFT-code \texttt{SeqQuest} \cite{seq} with
norm-conserving pseudopotentials and the generalized-gradient approximation of Perdew, Wang and Ernzerhof (PBE) \cite{PBE}
for exchange and correlation. Spin optimization is performed during all geometry relaxations.
The energy convergence criterion for all calculations is $1\times10^{-5}$ eV.
For smearing, we used the Gaussian method, together with a narrow width of smearing (0.005 eV) to make sure that the correct spin 
polarization can be achieved.
Diffusion pathways are investigated using the nudge elastic band (NEB) method \cite{NEB}
(as implemented in the \texttt{SeqQuest} code).
A 24$\times$24$\times$5 Monkhorst-Pack $k$-point mesh for the $1\times1\times1$ unit cell is considered and 
for larger supercells, it is adjusted accordingly. For all  calculations, grid spacing for the charge density integration is set to 0.16~$\AA$. 
Convergence tests with respect to k-point sampling and grid spacing show that calculated stresses are well converged (less than 0.0002 GPa).
For the charge density difference calculation, the relaxed structures from \texttt{SeqQuest} calculations were used in the DFT-code \texttt{VASP} \cite{vasp}. 
The unit cells for all \texttt{VASP} calculations were optimized again while the energy-forces criteria and spin polarization were chosen similar to ones used in \texttt{SeqQuest} calculations.

\begin{figure}
	\includegraphics[width=0.49\textwidth]{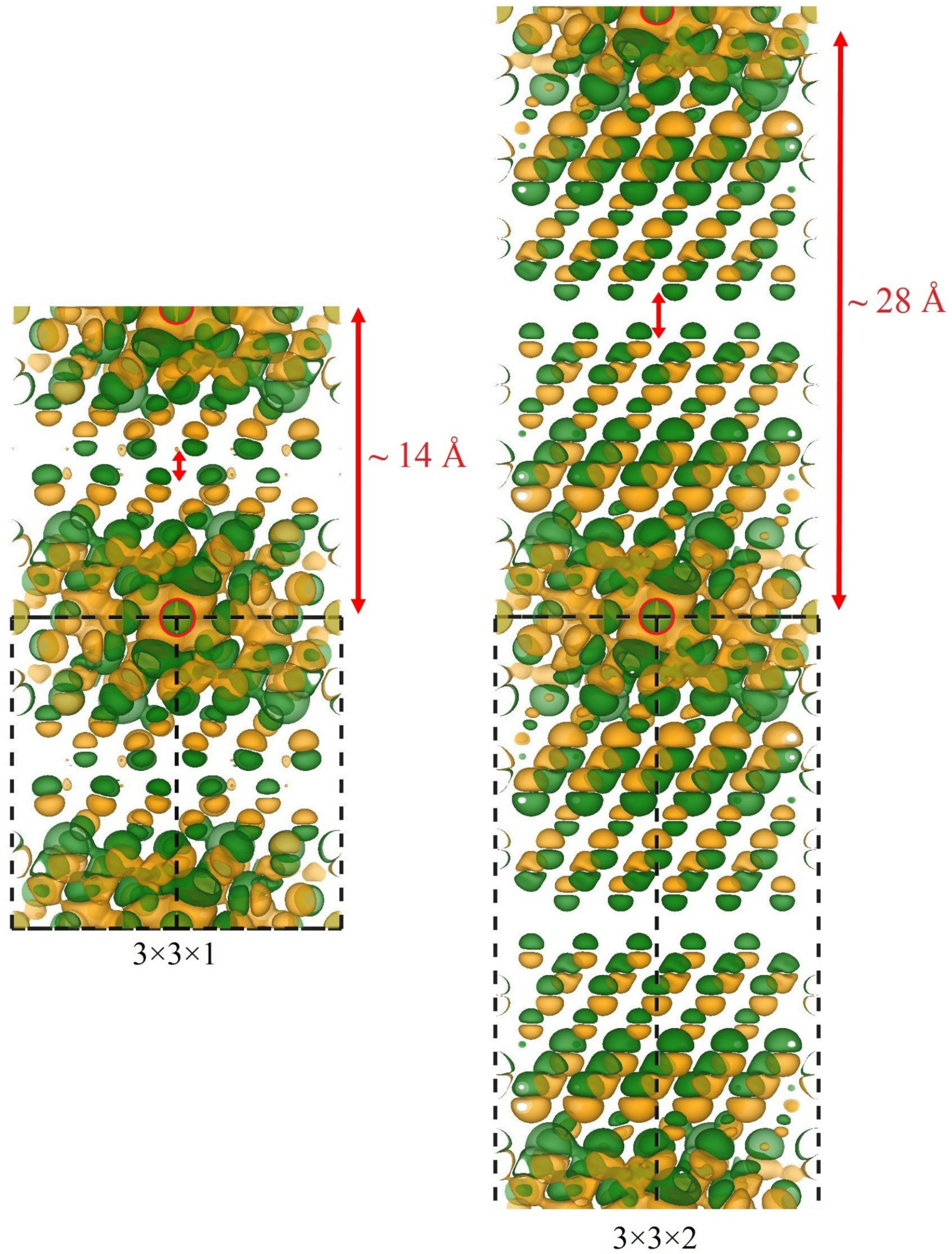}
	\caption{Charge density difference ($\rho_{\rm defective}-\rho_{\rm pristine}$) for 3$\times$3$\times$1 and 3$\times$3$\times$2 supercells.
		Green and yellow colors show charge accumulation and depletion, respectively, with an isosurface value of 0.001 $|e|/{\AA}^3$ for both supercells. The position of Li vacancy in the middle and orbitals separation are indicated with red circles and arrows, respectively. The supercells boundaries are also shown with dashed black lines.}
	\label{Fig-charge}
\end{figure}

\section{Results and discussion}
\subsection{I. Size dependence of elastic dipole tensor \\ ($G_{ij}$ in dilute limit)}

In order to capture finite-size effects, we used supercells with different volumes and aspect ratios. We also investigated the degree of coupling between lateral and longitudinal components by studying  various cell geometries. 
Figure~\ref{lateral} (left) shows the diagonal elements of the calculated stresses for isotropically repeated $n\times n\times n$ ($ 1 \leq n \leq 3 $) supercells containing a single (neutral) Li vacancy and  those of non-isotropic replicas $n\times n\times 1$ ($ 1 \leq n \leq 5 $, middle) and
$n\times n\times 2$ ($ 1 \leq n \leq 4 $, right) together with the second-order fit of the stresses and the diagonal components of the elastic dipole tensor. For the isotropic case (Fig. \ref{lateral} - left), due to computational limitations, the $4\times 4\times 4$ supercell is not calculated.   
In all cases, the non-diagonal elements exhibit small values. This is  due to the  fact that a Jahn--Teller distortion of MO$_2$ octahedrals occurs \cite{ouyang2,chen}, which can affect these non-diagonal elements of the elastic dipole tensor. The separation of defects along the $z$-axis between periodic images for $1\times 1 \times 1$ is equal to 13.96 \AA, while across the $xy$-plane, it is equal to 2.85 \AA. Therefore, we expect a stronger defect-defect interaction between the periodic images in the $xy$-direction rather than along the $z$-axis. 


In all geometries, where the $z$-extension was varied (see Fig. \ref{longitudinal}), the $xx$ and $yy$ components of the elastic dipole tensor are comparable (about 1.2 eV), while the $zz$ component is strongly varying and  is even  changing its sign in case of the $1\times1\times n$ cells  indicating a strong coupling between $xx$, $yy$ and $zz$ components.

We determined the components of the elastic dipole tensor by fitting the relation $\Delta \sigma_{ij}=-G_{ij}/V+\beta/V^2$ to the calculated data. The second term is accounting for the fact that higher order contributions to the elastic dipole tensor might be significant.
Using the mechanical definition of stress, we describe outward stress on the cell boundaries with positive and inward stress with negative signs. This means that the absence of positive (outward) stress would lead to cell contraction and vice versa ($\Delta V= \dfrac{1}{3B} {\rm Tr} \lbrace G_{ij} \rbrace$).

All data provide evidence for positive components of the dipole tensor in $x$- and $y$-direction and negative components in the $z$-direction. Therefore, the presence of a vacancy (during the deintercalation process) leads to a contraction in the $xy$-plane and expansion in the $z$-direction. These results are in agreement with the data reported by Xiong $et$ $al.$ \cite{ouyang2}.
The results for the $1\times1\times1$ cells deviate from the expected scaling behavior, since due to the high defect concentration, non-linear contributions prevail. Thus, these data are not shown in the plots. It is also evident that the data for the non-isotropically replicated supercells exhibit significant non-linear contributions for the $xx$- and $yy$-components of the defect elastic dipole tensor.

In Figure \ref{histo} we compare the values of the diagonal components of $G_{ij}$ using histogram plots. These plots show the scaling behavior from lateral (or longitudinal) extension alone towards isotropic one. It can be seen that the $G_{xx}$/$G_{yy}$ and $G_{zz}$ components do not show a linear scaling behavior. This is due to the fact that in these relatively small cells image-image interactions are still strong (especially across the $xy$-plane). 
 
Figure~\ref{Fig-charge} reveals the origin of the strong coupling between G$_{xx}$-G$_{yy}$ and G$_{zz}$ components. 
For this, we have plotted the charge density differences ($\rho_{\rm defective}-\rho_{\rm pristine}$) for two supercell dimensions, namely 3$\times$3$\times$1 and 3$\times$3$\times$2. 
For both supercells, similar isosurface value of 0.001 $|e|/{\AA}^3$ is considered. Figure~\ref{Fig-charge} indicates the charge distribution difference between defective and pristine structures as the lattice is increasing along $z$-axis. It can be seen that by lattice extension along the $z$-axis, the overlap of orbitals is decreasing which leads to the minimization of defect-defect interactions. However, due to the especial geometry of LiCoO$_2$ unit cell, despite the 3 times repeated cell along the $xy$-plane, the two defects between nearby images still show significant electrostatic interactions (although the overlap of orbitals along the $xy$-plane is also decreasing but with a less progress) which can be translated to the rather strong coupling of $xy$-plane stress components.   

To sum up this part, our results reveal that size and geometry effects have a massive influence on the calculation of the elastic dipole tensor because of the coupling between stress components. We see that the fits to the data scale more linearly as the cell sizes are increasing homogeneously ($n\times n\times n$). Therefore, non-linear terms affect the result the least for the isotropically scaled cell. 

\subsection{II. Defect formation energy under strain}

\begin{figure}
	\includegraphics[width=0.45\textwidth]{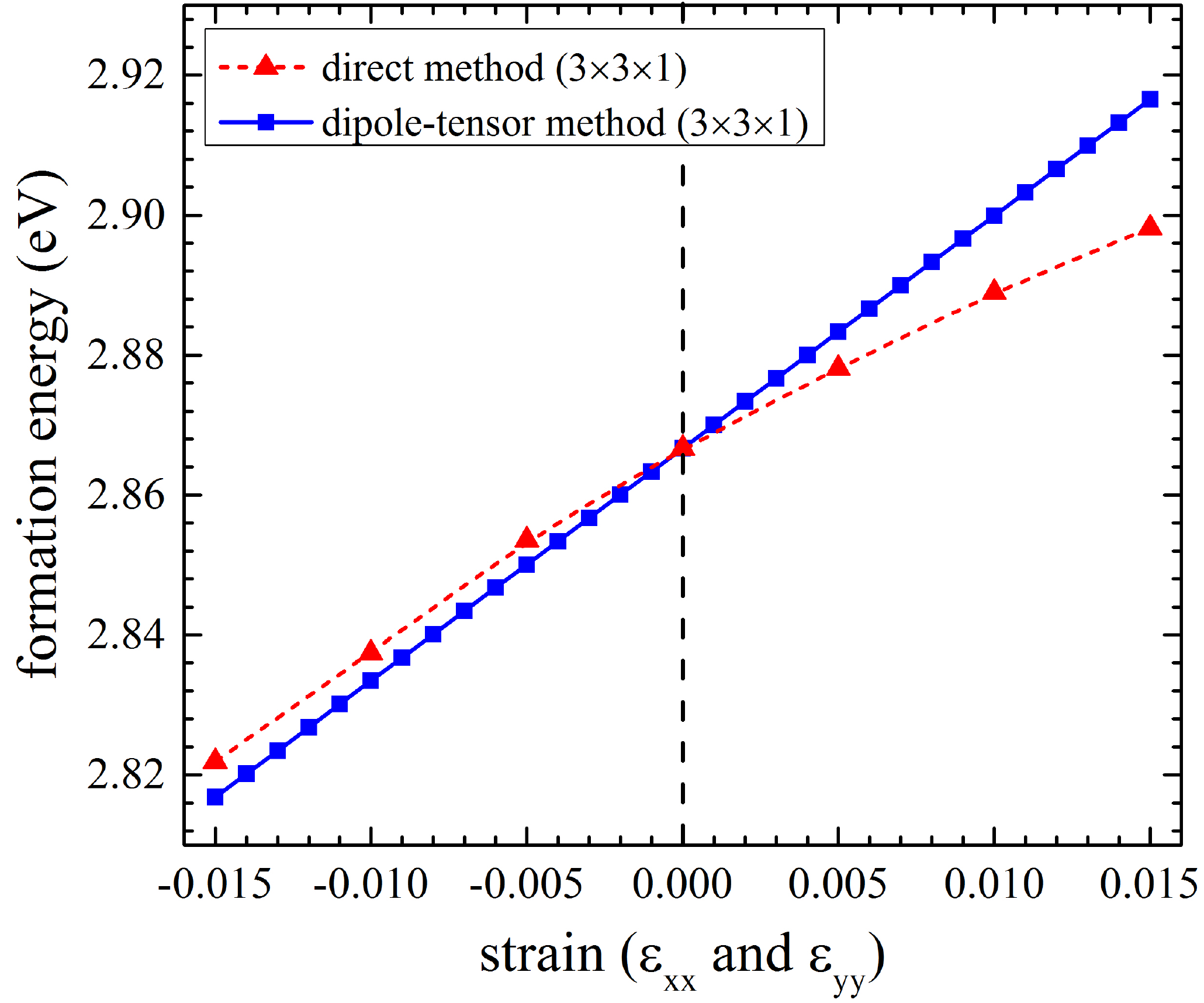}
	\includegraphics[width=0.45\textwidth]{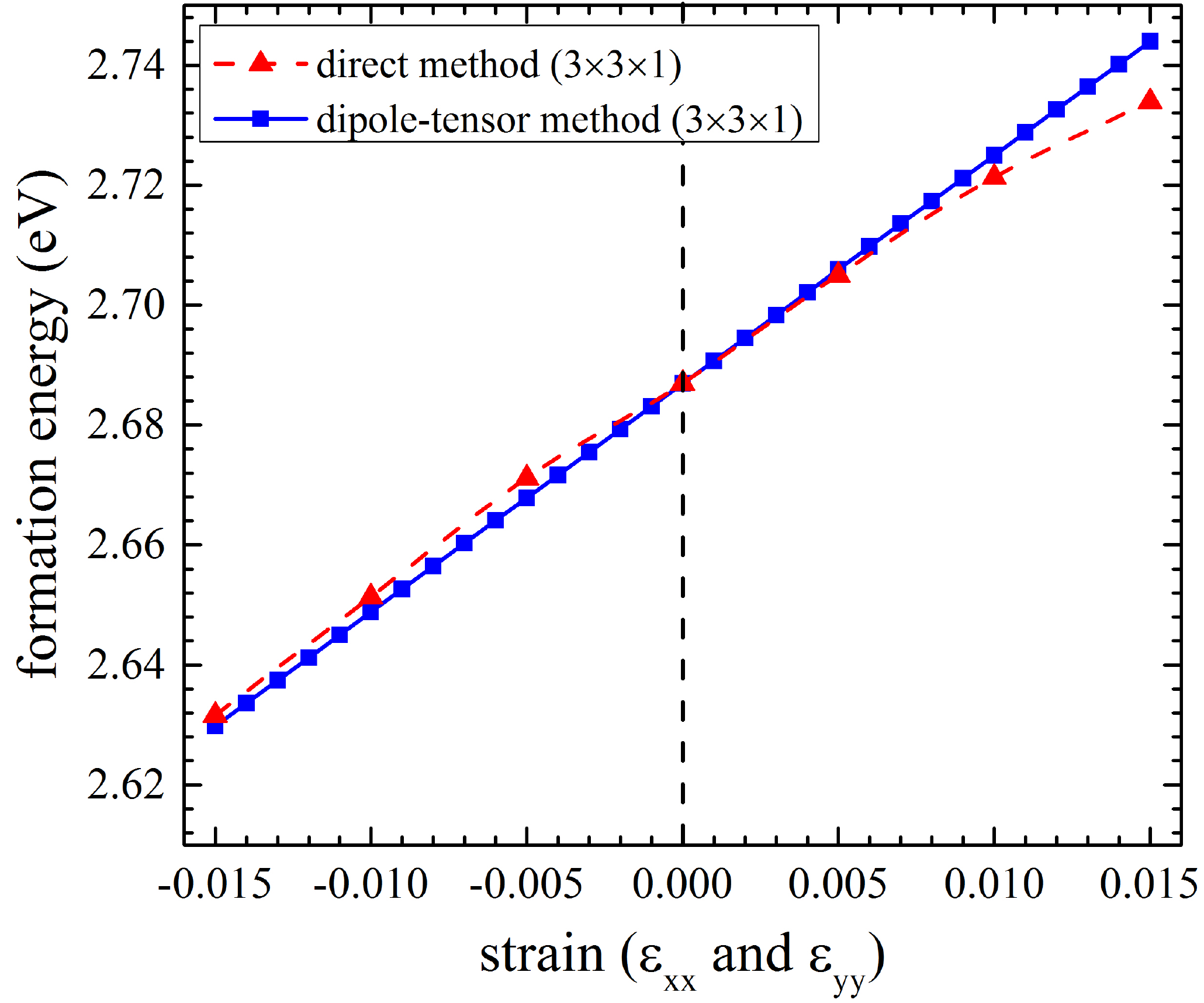}
	\caption{(color online) Top: \texttt{SeqQuest}-based calculation of defect formation energy as a function of strain (laterally, in the $xy-$plane) using direct and dipole-tensor methods 
		for a neutral single Li-vacancy in a 3$\times$3$\times$1 supercell. Bottom: \texttt{VASP}-based calculation of defect formation energy as a function of strain (in the $xy$-plane) using direct and dipole-tensor methods 
		for a neutral single Li-vacancy in a 3$\times$3$\times$1 supercell. By comparison between two plots, negligible difference between two methods of DFT calculations (atomic orbitals and plane waves basis sets and different pseudopotentials) is evident.}
	\label{form}
\end{figure}

In order to investigate the effect of strain on the defect formation energy in LiCoO$_2$, we used Eqs. \ref{eq:Ed_indirect} (dipole-tensor method) and \ref{eq:Ed_direct} (direct method). 
With \texttt{SeqQuest} we calculated a value of 2.86~ eV for the formation energy of a neutral Li vacancy in the 3$\times$3$\times$1 supercell as our reference for the unstrained case, which is in 
agreement with data previously reported by Hoang~$et$ $al.$ \cite{khang3}. 

Fig. \ref{form} shows the change in defect formation energy as a function of external strain (laterally, in the $xy-$plane) with two methods and two DFT codes. For all cases, the red points are obtained from the direct calculation in which 
the total energies of pristine and defective are strained. In this case, we can directly calculate the formation energies under strain. Since we need to calculate the total energies in each strain regime, this is computationally time-consuming. The second method is shown with the blue points which are obtained from Eq. \ref{eq:Ed_indirect}. In this case, we only need the total energies of unstrained pristine and defective supercells, together with the already-calculated $G_{ij}$ for that specific supercell.
 
The directly calculated strain-dependent defect formation energies show a slightly non-linear behavior and do not follow the prediction of linear elastic theory. Moreover, this non-linear behavior is not symmetric in case of 3$\times$3$\times$1 supercell using either DFT methods, as can be seen in Fig. \ref{form}. 

A similar trend for other systems has also been previously reported. For example Zhu $et \, al.$ investigated the effect of strain on the formation energy of Cu vacancy in Cu$_2$ZnSn(S,Se)$_4$ system \cite{Zhu2014} in which they show how symmetry breaking will lead to this deviation from linear elasticity. Similarly Aschauer $et \, al.$ found a non-linear behavior for the strain dependence of oxygen formation in MnO \cite{Aschauer2015}. The reason why linear elastic theory fails to predict this trend is first due to the fact that there are higher order terms relevant in the Taylor expansion given in Eq. \ref{eq:energy_1}. Another reason is that the elastic constants of the defective system are not the same as in pristine system. 
Regardless of the methods and supercell sizes, it can also be concluded that under the influence of lateral elastic strain, the formation energy of a single neutral vacancy in bulk LiCoO$_2$ varies by about 0.02 eV (with $\epsilon=1 \%$).    

\subsection{III. Migration barrier of Li vacancy under strain}

\begin{figure}[!th]
	\includegraphics[width=0.45\textwidth]{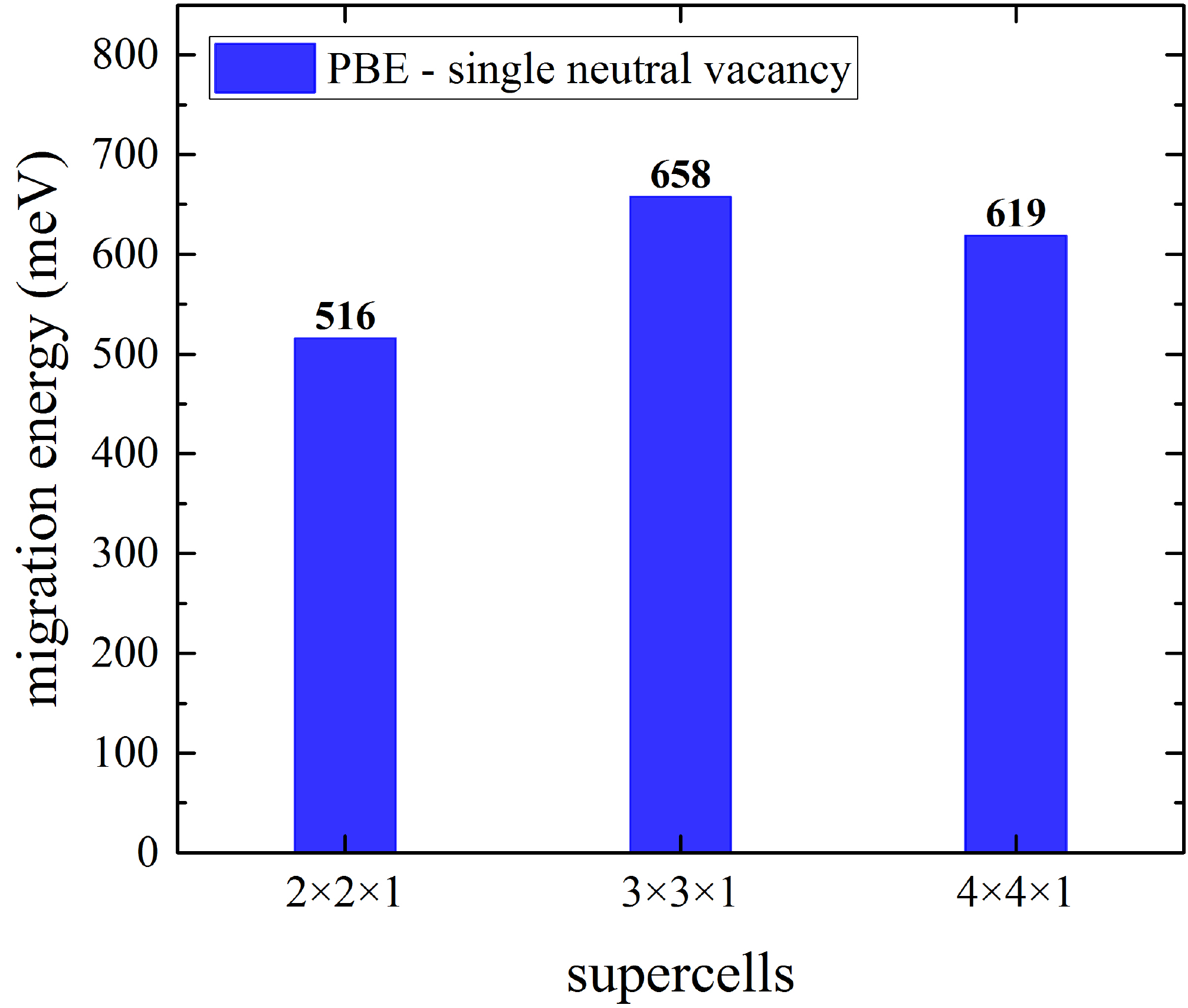}
	\caption{(color online) Calculated migration energy barriers of neutral Li vacancy jump in bulk LiCoO$_2$. Given are the results for different supercell sizes.}
	\label{histo-mig}
\end{figure}

\begin{figure}[!th]
	\includegraphics[width=0.45\textwidth]{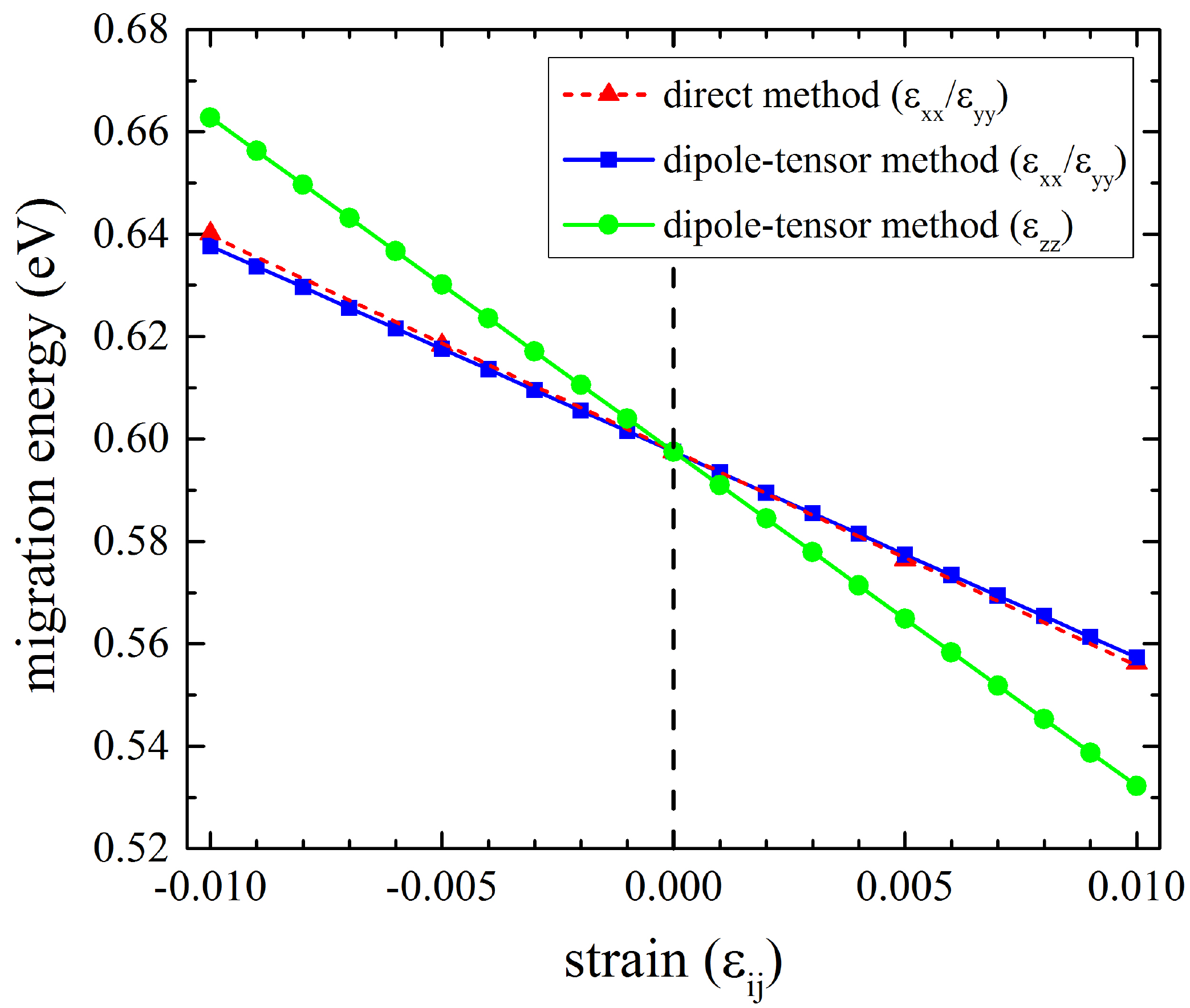}
	\caption{(color online) Diffusion energy barrier vs. applied external strain (laterally, in the $ab-$plane) using direct and dipole-tensor methods 
		for a neutral single Li-vacancy in a 3$\times$3$\times$1 supercell. The red points correspond to NEB results in which a plane strain regime is applied. A perfect agreement with the dipole-tensor method (blue points-line) for the similar strain regime can be seen. Green points-line shows the dipole tensor method for the longitudinal strain (along $c-$axis) which indicates a stronger effect compared to strain in the $ab-$plane.}
	\label{mig}
\end{figure}

We now focus on the effect of strain on the Li diffusion for the pathway indicated in Figs. \ref{str} (b) and (c). 
In the first section of results and discussion, we calculated the elastic dipole tensor in the dilute limit and revealed that size and geometry of the supercell have a significant influence.
This is why we first investigated the influence of cell size on the migration barrier for Li vacancy jumps using the NEB method. The results are shown in Fig. \ref{histo-mig}. The calculated values for the 3$\times$3$\times$1 and 4$\times$4$\times$1 deviate by about 10\%, while the smaller cell shows a significantly smaller migration barrier. 
In order to reduced computational efforts, we have chosen the 3$\times$3$\times$1 supercell in order to investigate the strain effect on migration energy barrier. 

Figure \ref{mig} shows migration barriers for vacancy hopping as function of strain using the directly calculated data (red in the $xy$-plane) and those obtained from the elastic dipole tensor method (blue in the $xy$-plane and green along the $z$-axis) for 3$\times$3$\times$1 supercell using Eq. \ref{eq13}. For the direct calculations, four lateral strain states are introduced homogeneously on the lattice parameters of $a$ and $b$. 
The calculation of the elastic dipole tensor in Eq. \ref{eq13} was performed at the initial and saddle points for the 3$\times$3$\times$1 supercell. 
Since we already calculated $G_{ij}$ for the initial state (Fig. \ref{lateral}), we used the geometry of the saddle point from the NEB calculation for the unstrained case to obtain the $G_{ij}$ for the transition state. We obtain the following values for the elastic dipole tensor in the initial state and at the saddle point: \\

$ G_{\rm initial}=\begin{bmatrix}    
\ 1.66 & 0.01 & 0.01 \\
0.01 & 1.65 &  -0.02 \\
0.02 & -0.02 & -1.72 \\
\end{bmatrix}$\\

$ G_{\rm saddle}=\begin{bmatrix}    
\ -0.08 & -0.81 & 1.48 \\
-0.81 & 0.77 & -2.30 \\
1.60 & -2.49 & 8.24 \\
\end{bmatrix}$\\

By considering the signs of diagonal components of elastic dipole tensor for the initial state, the direction of the gray arrows in Figs. \ref{str}-b and \ref{str}-c are justified.   

Fig. \ref{mig} shows the very good agreement between the directly calculated data and those obtained from elastic-dipole tensor in the $xy$-plane. Due to computational limitations, we do not compare the two methods for the longitudinal stain, however, we postulate that the good agreement observed for the $xy$-plane strain is also established for the $z$-axis strain. 
It can also be seen that all curves behave linearly but with different slopes between lateral and longitudinal strains (between red-blue and green plots). By applying a positive strain, the energy barrier decreases and vice versa, which is also in agreement with  previous results from Ning et. al. \cite{ning}. It can be seen that, as a result of tensile strain, there is more Li-Li separation and less Li-Li repulsion and therefore, moving Li atom can intercalate/deintercalate more easily. From another point of view, applying strain could disturb the octahedrals orientation and therefore affects the 
potential energy surface which directly influences the energy barrier. 

Since we see a good agreement between two methods in Fig. \ref{mig} for the lateral strain, we also plotted the effect of longitudinal strain on the migration energy barrier which is indicated with green color. It is clear from Fig. \ref{mig} that the longitudinal strain has a much stronger effect on the energy barrier, compared to lateral strain. 
Therefore, this shows that instead of significant computational efforts due to performing NEB calculations at each strain regime, it is possible to use Eq. \ref{eq13} and obtain the same results with much less computational resources.  

A comparison between Figs. \ref{form} and \ref{mig} also indicates that the effect of lateral ($xy$-plane) 
strain on the Li vacancy formation and migration are opposite of each other. This means that, while a lateral compressive strain decreases the energy barrier (due to above mentioned reasons), 
the same strain results in a larger Li vacancy formation energy. However, as can be seen from the plots, this effect on formation and migration energy is not equal. 
Therefore in the activation energy, which is the sum of formation and migration, at a lateral strain regime equals to $+1\%$, for example, the migration term is decreased by 0.04 eV while formation term is increased by 0.02 eV. Thus, at this strain value, the overall trend is 
a decrease in the activation energy by 0.02 eV. Considering the relation of diffusivity, 
\begin{equation}
\begin{aligned}
D\approx{\rm exp}\bigg({-\dfrac{\Delta G_{\rm A}}{k_BT}}\bigg), \nonumber
\end{aligned}
\end{equation}
decreasing of 0.02 eV in activation energy leads to almost five times increase in the diffusivity and hence the conductivity. Therefore, the massive strain effect on ionic conductivity in bulk LiCoO$_2$ is evident. We again note that the effect of longitudinal strain is even more dominant than the lateral one. Particularly, only 1\% strain along the $z$-axis can change the conductivity up to ten times compared to unstrained case. Moreover, according to theoretical findings \cite{Li}, 4-5\% and experimental evidences \cite{wang2,Takahashi}, 2-3\% strain is expected during the lithiation/delithiation of LiCoO$_2$ to Li$_{0.5}$CoO$_2$. Therefore, strain fields can significantly influence the ionic conductivity in bulk LiCoO$_2$ which shows that it can be employed to tune the particle mobility in battery materials.

\section{Summary and CONCLUSIONS}

In summary, we have performed a detailed comparison for the variations of defect formation and migration energies with respect to strain in bulk LiCO$_2$ using (i) direct evaluation of strained supercells (at each strain regime separately) and (ii) classical elasticity theory by computing the elastic dipole tensor ($G_{ij}$). The latter method requires only three DFT calculations (pristine and defected cells for formation energies plus defected cell in transition state for energy barriers) for evaluation of both formation and migration energies to obtain the full variations as a function of any strain state. 

We found that the calculated formation energies using the elastic dipole tensor method deviate slightly when they are compared with the direct method (by less then 1\% for $\epsilon < 0.015$). We note tha, however, deviation to some degree depends on particular basis sets and/or pseudopotentials. Moreover, the mentioned deviation largely cancels out for migration barriers and it makes this method even more error-free in case of strain-induced diffusion analysis. Therefore, using the elastic dipole tensor method for the analysis of strained-induced formation and migration energies is computationally very efficient. 

We also highlight that estimating the elastic dipole tensor in the dilute limit can be affected by  finite-size effects and coupling between the stress components.  
Moreover, we found that the contribution of migration energy to the total activation energy when a lateral strain regime is applied, is more dominant than the contribution of formation energy. 
Finally, we can conclude that the effect of even small strains on ionic transport properties in bulk LiCoO$_2$ is very significant. We showed that only 1\% in-plane strain can change the conductivity by a factor of 5, while the presence of out-of-plane strains can 
change the conductivity by an order of magnitude.


\begin{center}$\bf{Acknowledgments}$\end{center}
\noindent
AM and PK gratefully acknowledge support from the 
"Bundesministerium f\"ur Bildung und Forschung"
(BMBF), the computing time
granted on the Hessian high performance computer "LICHTENBERG"
and Zentraleinrichtung f\"ur Datenverarbeitung (ZEDAT) at the Freie
Universit\"at Berlin. JR and KA acknowledge support through SPP 1473 of the German Research Foundation and project DFG-578/19-1.


%


\begin{thebibliography}{50}%
\makeatletter
\providecommand \@ifxundefined [1]{%
 \@ifx{#1\undefined}
}%
\providecommand \@ifnum [1]{%
 \ifnum #1\expandafter \@firstoftwo
 \else \expandafter \@secondoftwo
 \fi
}%
\providecommand \@ifx [1]{%
 \ifx #1\expandafter \@firstoftwo
 \else \expandafter \@secondoftwo
 \fi
}%
\providecommand \natexlab [1]{#1}%
\providecommand \enquote  [1]{``#1''}%
\providecommand \bibnamefont  [1]{#1}%
\providecommand \bibfnamefont [1]{#1}%
\providecommand \citenamefont [1]{#1}%
\providecommand \href@noop [0]{\@secondoftwo}%
\providecommand \href [0]{\begingroup \@sanitize@url \@href}%
\providecommand \@href[1]{\@@startlink{#1}\@@href}%
\providecommand \@@href[1]{\endgroup#1\@@endlink}%
\providecommand \@sanitize@url [0]{\catcode `\\12\catcode `\$12\catcode
  `\&12\catcode `\#12\catcode `\^12\catcode `\_12\catcode `\%12\relax}%
\providecommand \@@startlink[1]{}%
\providecommand \@@endlink[0]{}%
\providecommand \url  [0]{\begingroup\@sanitize@url \@url }%
\providecommand \@url [1]{\endgroup\@href {#1}{\urlprefix }}%
\providecommand \urlprefix  [0]{URL }%
\providecommand \Eprint [0]{\href }%
\providecommand \doibase [0]{http://dx.doi.org/}%
\providecommand \selectlanguage [0]{\@gobble}%
\providecommand \bibinfo  [0]{\@secondoftwo}%
\providecommand \bibfield  [0]{\@secondoftwo}%
\providecommand \translation [1]{[#1]}%
\providecommand \BibitemOpen [0]{}%
\providecommand \bibitemStop [0]{}%
\providecommand \bibitemNoStop [0]{.\EOS\space}%
\providecommand \EOS [0]{\spacefactor3000\relax}%
\providecommand \BibitemShut  [1]{\csname bibitem#1\endcsname}%
\let\auto@bib@innerbib\@empty
\bibitem [{\citenamefont {Sun}\ \emph {et~al.}(2007)\citenamefont {Sun},
  \citenamefont {Thompson},\ and\ \citenamefont {Nishida}}]{Sun}%
  \BibitemOpen
  \bibfield  {author} {\bibinfo {author} {\bibfnamefont {Y.}~\bibnamefont
  {Sun}}, \bibinfo {author} {\bibfnamefont {S.~E.}\ \bibnamefont {Thompson}}, \
  and\ \bibinfo {author} {\bibfnamefont {T.}~\bibnamefont {Nishida}},\ }\href
  {\doibase 10.1063/1.2730561} {\bibfield  {journal} {\bibinfo  {journal}
  {Journal of Applied Physics}\ }\textbf {\bibinfo {volume} {101}},\ \bibinfo
  {pages} {104503} (\bibinfo {year} {2007})}\BibitemShut {NoStop}%
\bibitem [{\citenamefont {Penev}\ \emph {et~al.}(2001)\citenamefont {Penev},
  \citenamefont {Kratzer},\ and\ \citenamefont {Scheffler}}]{Penev}%
  \BibitemOpen
  \bibfield  {author} {\bibinfo {author} {\bibfnamefont {E.}~\bibnamefont
  {Penev}}, \bibinfo {author} {\bibfnamefont {P.}~\bibnamefont {Kratzer}}, \
  and\ \bibinfo {author} {\bibfnamefont {M.}~\bibnamefont {Scheffler}},\
  }\href@noop {} {\bibfield  {journal} {\bibinfo  {journal} {Physical Review
  B}\ }\textbf {\bibinfo {volume} {64}} (\bibinfo {year} {2001})}\BibitemShut
  {NoStop}%
\bibitem [{\citenamefont {Hubbard}\ \emph {et~al.}(2008)\citenamefont
  {Hubbard}, \citenamefont {Cress}, \citenamefont {Bailey}, \citenamefont
  {Raffaelle}, \citenamefont {Bailey},\ and\ \citenamefont {Wilt}}]{Hubbard}%
  \BibitemOpen
  \bibfield  {author} {\bibinfo {author} {\bibfnamefont {S.~M.}\ \bibnamefont
  {Hubbard}}, \bibinfo {author} {\bibfnamefont {C.~D.}\ \bibnamefont {Cress}},
  \bibinfo {author} {\bibfnamefont {C.~G.}\ \bibnamefont {Bailey}}, \bibinfo
  {author} {\bibfnamefont {R.~P.}\ \bibnamefont {Raffaelle}}, \bibinfo {author}
  {\bibfnamefont {S.~G.}\ \bibnamefont {Bailey}}, \ and\ \bibinfo {author}
  {\bibfnamefont {D.~M.}\ \bibnamefont {Wilt}},\ }\href {\doibase
  10.1063/1.2903699} {\bibfield  {journal} {\bibinfo  {journal} {Applied
  Physics Letters}\ }\textbf {\bibinfo {volume} {92}},\ \bibinfo {pages}
  {123512} (\bibinfo {year} {2008})}\BibitemShut {NoStop}%
\bibitem [{\citenamefont {Sheu}\ \emph {et~al.}(2005)\citenamefont {Sheu},
  \citenamefont {Yang}, \citenamefont {Wang}, \citenamefont {Chang},
  \citenamefont {Huang}, \citenamefont {Huang}, \citenamefont {Chen},\ and\
  \citenamefont {Diaz}}]{yi}%
  \BibitemOpen
  \bibfield  {author} {\bibinfo {author} {\bibfnamefont {Y.-M.}\ \bibnamefont
  {Sheu}}, \bibinfo {author} {\bibfnamefont {S.-J.}\ \bibnamefont {Yang}},
  \bibinfo {author} {\bibfnamefont {C.-C.}\ \bibnamefont {Wang}}, \bibinfo
  {author} {\bibfnamefont {C.-S.}\ \bibnamefont {Chang}}, \bibinfo {author}
  {\bibfnamefont {L.-P.}\ \bibnamefont {Huang}}, \bibinfo {author}
  {\bibfnamefont {T.-Y.}\ \bibnamefont {Huang}}, \bibinfo {author}
  {\bibfnamefont {M.-J.}\ \bibnamefont {Chen}}, \ and\ \bibinfo {author}
  {\bibfnamefont {C.}~\bibnamefont {Diaz}},\ }\href {\doibase
  10.1109/ted.2004.841286} {\bibfield  {journal} {\bibinfo  {journal} {{IEEE}
  Trans. Electron Devices}\ }\textbf {\bibinfo {volume} {52}},\ \bibinfo
  {pages} {30} (\bibinfo {year} {2005})}\BibitemShut {NoStop}%
\bibitem [{\citenamefont {Zhang}\ \emph {et~al.}(2011)\citenamefont {Zhang},
  \citenamefont {Cui},\ and\ \citenamefont {Wang}}]{cui}%
  \BibitemOpen
  \bibfield  {author} {\bibinfo {author} {\bibfnamefont {Q.}~\bibnamefont
  {Zhang}}, \bibinfo {author} {\bibfnamefont {Y.}~\bibnamefont {Cui}}, \ and\
  \bibinfo {author} {\bibfnamefont {E.}~\bibnamefont {Wang}},\ }\href {\doibase
  10.1021/jp1115977} {\bibfield  {journal} {\bibinfo  {journal} {J. Phys. Chem.
  C}\ }\textbf {\bibinfo {volume} {115}},\ \bibinfo {pages} {9376} (\bibinfo
  {year} {2011})}\BibitemShut {NoStop}%
\bibitem [{\citenamefont {Deshpande}\ \emph {et~al.}(2010)\citenamefont
  {Deshpande}, \citenamefont {Cheng},\ and\ \citenamefont
  {Verbrugge}}]{rutooj}%
  \BibitemOpen
  \bibfield  {author} {\bibinfo {author} {\bibfnamefont {R.}~\bibnamefont
  {Deshpande}}, \bibinfo {author} {\bibfnamefont {Y.-T.}\ \bibnamefont
  {Cheng}}, \ and\ \bibinfo {author} {\bibfnamefont {M.~W.}\ \bibnamefont
  {Verbrugge}},\ }\href {\doibase 10.1016/j.jpowsour.2010.02.021} {\bibfield
  {journal} {\bibinfo  {journal} {Journal of Power Sources}\ }\textbf {\bibinfo
  {volume} {195}},\ \bibinfo {pages} {5081} (\bibinfo {year}
  {2010})}\BibitemShut {NoStop}%
\bibitem [{\citenamefont {Tealdi}\ \emph {et~al.}(2016)\citenamefont {Tealdi},
  \citenamefont {Heath},\ and\ \citenamefont {Islam}}]{seif}%
  \BibitemOpen
  \bibfield  {author} {\bibinfo {author} {\bibfnamefont {C.}~\bibnamefont
  {Tealdi}}, \bibinfo {author} {\bibfnamefont {J.}~\bibnamefont {Heath}}, \
  and\ \bibinfo {author} {\bibfnamefont {M.~S.}\ \bibnamefont {Islam}},\ }\href
  {\doibase 10.1039/c5ta09418f} {\bibfield  {journal} {\bibinfo  {journal} {J.
  Mater. Chem. A}\ }\textbf {\bibinfo {volume} {4}},\ \bibinfo {pages} {6998}
  (\bibinfo {year} {2016})}\BibitemShut {NoStop}%
\bibitem [{\citenamefont {Malik}\ \emph {et~al.}(2010)\citenamefont {Malik},
  \citenamefont {Burch}, \citenamefont {Bazant},\ and\ \citenamefont
  {Ceder}}]{malik}%
  \BibitemOpen
  \bibfield  {author} {\bibinfo {author} {\bibfnamefont {R.}~\bibnamefont
  {Malik}}, \bibinfo {author} {\bibfnamefont {D.}~\bibnamefont {Burch}},
  \bibinfo {author} {\bibfnamefont {M.}~\bibnamefont {Bazant}}, \ and\ \bibinfo
  {author} {\bibfnamefont {G.}~\bibnamefont {Ceder}},\ }\href {\doibase
  10.1021/nl1023595} {\bibfield  {journal} {\bibinfo  {journal} {Nano Letters}\
  }\textbf {\bibinfo {volume} {10}},\ \bibinfo {pages} {4123} (\bibinfo {year}
  {2010})}\BibitemShut {NoStop}%
\bibitem [{\citenamefont {Wang}\ \emph {et~al.}(2007)\citenamefont {Wang},
  \citenamefont {Sone}, \citenamefont {Segami}, \citenamefont {Naito},
  \citenamefont {Yamada},\ and\ \citenamefont {Kibe}}]{wang}%
  \BibitemOpen
  \bibfield  {author} {\bibinfo {author} {\bibfnamefont {X.}~\bibnamefont
  {Wang}}, \bibinfo {author} {\bibfnamefont {Y.}~\bibnamefont {Sone}}, \bibinfo
  {author} {\bibfnamefont {G.}~\bibnamefont {Segami}}, \bibinfo {author}
  {\bibfnamefont {H.}~\bibnamefont {Naito}}, \bibinfo {author} {\bibfnamefont
  {C.}~\bibnamefont {Yamada}}, \ and\ \bibinfo {author} {\bibfnamefont
  {K.}~\bibnamefont {Kibe}},\ }\href {\doibase 10.1149/1.2386933} {\bibfield
  {journal} {\bibinfo  {journal} {Journal of The Electrochemical Society}\
  }\textbf {\bibinfo {volume} {154}},\ \bibinfo {pages} {A14} (\bibinfo {year}
  {2007})}\BibitemShut {NoStop}%
\bibitem [{\citenamefont {Diehm}\ \emph {et~al.}(2012)\citenamefont {Diehm},
  \citenamefont {{\'{A}}goston},\ and\ \citenamefont {Albe}}]{karsten3}%
  \BibitemOpen
  \bibfield  {author} {\bibinfo {author} {\bibfnamefont {P.~M.}\ \bibnamefont
  {Diehm}}, \bibinfo {author} {\bibfnamefont {P.}~\bibnamefont
  {{\'{A}}goston}}, \ and\ \bibinfo {author} {\bibfnamefont {K.}~\bibnamefont
  {Albe}},\ }\href {\doibase 10.1002/cphc.201200257} {\bibfield  {journal}
  {\bibinfo  {journal} {{ChemPhysChem}}\ }\textbf {\bibinfo {volume} {13}},\
  \bibinfo {pages} {2443} (\bibinfo {year} {2012})}\BibitemShut {NoStop}%
\bibitem [{\citenamefont {Walukiewicz}(1994)}]{Walu}%
  \BibitemOpen
  \bibfield  {author} {\bibinfo {author} {\bibfnamefont {W.}~\bibnamefont
  {Walukiewicz}},\ }\href {\doibase 10.1103/physrevb.50.5221} {\bibfield
  {journal} {\bibinfo  {journal} {Phys. Rev. B}\ }\textbf {\bibinfo {volume}
  {50}},\ \bibinfo {pages} {5221} (\bibinfo {year} {1994})}\BibitemShut
  {NoStop}%
\bibitem [{\citenamefont {Azpiroz}\ \emph {et~al.}(2015)\citenamefont
  {Azpiroz}, \citenamefont {Mosconi}, \citenamefont {Bisquert},\ and\
  \citenamefont {Angelis}}]{Azp}%
  \BibitemOpen
  \bibfield  {author} {\bibinfo {author} {\bibfnamefont {J.~M.}\ \bibnamefont
  {Azpiroz}}, \bibinfo {author} {\bibfnamefont {E.}~\bibnamefont {Mosconi}},
  \bibinfo {author} {\bibfnamefont {J.}~\bibnamefont {Bisquert}}, \ and\
  \bibinfo {author} {\bibfnamefont {F.~D.}\ \bibnamefont {Angelis}},\ }\href
  {\doibase 10.1039/c5ee01265a} {\bibfield  {journal} {\bibinfo  {journal}
  {Energy Environ. Sci.}\ }\textbf {\bibinfo {volume} {8}},\ \bibinfo {pages}
  {2118} (\bibinfo {year} {2015})}\BibitemShut {NoStop}%
\bibitem [{\citenamefont {Fisher}\ \emph {et~al.}(2008)\citenamefont {Fisher},
  \citenamefont {Prieto},\ and\ \citenamefont {Islam}}]{Fisher}%
  \BibitemOpen
  \bibfield  {author} {\bibinfo {author} {\bibfnamefont {C.~A.~J.}\
  \bibnamefont {Fisher}}, \bibinfo {author} {\bibfnamefont {V.~M.~H.}\
  \bibnamefont {Prieto}}, \ and\ \bibinfo {author} {\bibfnamefont {M.~S.}\
  \bibnamefont {Islam}},\ }\href {\doibase 10.1021/cm801262x} {\bibfield
  {journal} {\bibinfo  {journal} {Chemistry of Materials}\ }\textbf {\bibinfo
  {volume} {20}},\ \bibinfo {pages} {5907} (\bibinfo {year}
  {2008})}\BibitemShut {NoStop}%
\bibitem [{\citenamefont {Freedman}\ \emph {et~al.}(2009)\citenamefont
  {Freedman}, \citenamefont {Roundy},\ and\ \citenamefont {Arias}}]{Freedman}%
  \BibitemOpen
  \bibfield  {author} {\bibinfo {author} {\bibfnamefont {D.~A.}\ \bibnamefont
  {Freedman}}, \bibinfo {author} {\bibfnamefont {D.}~\bibnamefont {Roundy}}, \
  and\ \bibinfo {author} {\bibfnamefont {T.~A.}\ \bibnamefont {Arias}},\ }\href
  {\doibase 10.1103/physrevb.80.064108} {\bibfield  {journal} {\bibinfo
  {journal} {Phys. Rev. B}\ }\textbf {\bibinfo {volume} {80}} (\bibinfo {year}
  {2009}),\ 10.1103/physrevb.80.064108}\BibitemShut {NoStop}%
\bibitem [{\citenamefont {Goyal}\ \emph {et~al.}(2015)\citenamefont {Goyal},
  \citenamefont {Phillpot}, \citenamefont {Subramanian}, \citenamefont
  {Andersson}, \citenamefont {Stanek},\ and\ \citenamefont {Uberuaga}}]{Goyal}%
  \BibitemOpen
  \bibfield  {author} {\bibinfo {author} {\bibfnamefont {A.}~\bibnamefont
  {Goyal}}, \bibinfo {author} {\bibfnamefont {S.~R.}\ \bibnamefont {Phillpot}},
  \bibinfo {author} {\bibfnamefont {G.}~\bibnamefont {Subramanian}}, \bibinfo
  {author} {\bibfnamefont {D.~A.}\ \bibnamefont {Andersson}}, \bibinfo {author}
  {\bibfnamefont {C.~R.}\ \bibnamefont {Stanek}}, \ and\ \bibinfo {author}
  {\bibfnamefont {B.~P.}\ \bibnamefont {Uberuaga}},\ }\href {\doibase
  10.1103/physrevb.91.094103} {\bibfield  {journal} {\bibinfo  {journal} {Phys.
  Rev. B}\ }\textbf {\bibinfo {volume} {91}} (\bibinfo {year} {2015}),\
  10.1103/physrevb.91.094103}\BibitemShut {NoStop}%
\bibitem [{\citenamefont {Moradabadi}\ and\ \citenamefont
  {Kaghazchi}(2015)}]{ashkan}%
  \BibitemOpen
  \bibfield  {author} {\bibinfo {author} {\bibfnamefont {A.}~\bibnamefont
  {Moradabadi}}\ and\ \bibinfo {author} {\bibfnamefont {P.}~\bibnamefont
  {Kaghazchi}},\ }\href {\doibase 10.1039/c5cp02246k} {\ \textbf {\bibinfo
  {volume} {17}},\ \bibinfo {pages} {22917} (\bibinfo {year}
  {2015})}\BibitemShut {NoStop}%
\bibitem [{\citenamefont {Moradabadi}\ and\ \citenamefont
  {Kaghazchi}(2016)}]{ashkan2}%
  \BibitemOpen
  \bibfield  {author} {\bibinfo {author} {\bibfnamefont {A.}~\bibnamefont
  {Moradabadi}}\ and\ \bibinfo {author} {\bibfnamefont {P.}~\bibnamefont
  {Kaghazchi}},\ }\href {\doibase 10.1063/1.4952434} {\bibfield  {journal}
  {\bibinfo  {journal} {Appl. Phys. Lett.}\ }\textbf {\bibinfo {volume}
  {108}},\ \bibinfo {pages} {213906} (\bibinfo {year} {2016})}\BibitemShut
  {NoStop}%
\bibitem [{\citenamefont {Hoang}\ and\ \citenamefont
  {Johannes}(2016)}]{khang1}%
  \BibitemOpen
  \bibfield  {author} {\bibinfo {author} {\bibfnamefont {K.}~\bibnamefont
  {Hoang}}\ and\ \bibinfo {author} {\bibfnamefont {M.}~\bibnamefont
  {Johannes}},\ }\href {\doibase 10.1021/acs.chemmater.5b04219} {\bibfield
  {journal} {\bibinfo  {journal} {Chemistry of Materials}\ }\textbf {\bibinfo
  {volume} {28}},\ \bibinfo {pages} {1325} (\bibinfo {year}
  {2016})}\BibitemShut {NoStop}%
\bibitem [{\citenamefont {Hoang}\ and\ \citenamefont
  {Johannes}(2012)}]{khang2}%
  \BibitemOpen
  \bibfield  {author} {\bibinfo {author} {\bibfnamefont {K.}~\bibnamefont
  {Hoang}}\ and\ \bibinfo {author} {\bibfnamefont {M.~D.}\ \bibnamefont
  {Johannes}},\ }\href {\doibase 10.1016/j.jpowsour.2012.01.126} {\bibfield
  {journal} {\bibinfo  {journal} {Journal of Power Sources}\ }\textbf {\bibinfo
  {volume} {206}},\ \bibinfo {pages} {274} (\bibinfo {year}
  {2012})}\BibitemShut {NoStop}%
\bibitem [{\citenamefont {Hoang}\ and\ \citenamefont
  {Johannes}(2014)}]{khang3}%
  \BibitemOpen
  \bibfield  {author} {\bibinfo {author} {\bibfnamefont {K.}~\bibnamefont
  {Hoang}}\ and\ \bibinfo {author} {\bibfnamefont {M.~D.}\ \bibnamefont
  {Johannes}},\ }\href {\doibase 10.1039/c4ta00673a} {\bibfield  {journal}
  {\bibinfo  {journal} {J. Mater. Chem. A}\ }\textbf {\bibinfo {volume} {2}},\
  \bibinfo {pages} {5224} (\bibinfo {year} {2014})}\BibitemShut {NoStop}%
\bibitem [{\citenamefont {Guo}\ \emph {et~al.}(2016)\citenamefont {Guo},
  \citenamefont {Song}, \citenamefont {Zhuo}, \citenamefont {Hu}, \citenamefont
  {Liu}, \citenamefont {Duan}, \citenamefont {Zheng}, \citenamefont {Chen},
  \citenamefont {Yang}, \citenamefont {Amine},\ and\ \citenamefont
  {Pan}}]{amine}%
  \BibitemOpen
  \bibfield  {author} {\bibinfo {author} {\bibfnamefont {H.}~\bibnamefont
  {Guo}}, \bibinfo {author} {\bibfnamefont {X.}~\bibnamefont {Song}}, \bibinfo
  {author} {\bibfnamefont {Z.}~\bibnamefont {Zhuo}}, \bibinfo {author}
  {\bibfnamefont {J.}~\bibnamefont {Hu}}, \bibinfo {author} {\bibfnamefont
  {T.}~\bibnamefont {Liu}}, \bibinfo {author} {\bibfnamefont {Y.}~\bibnamefont
  {Duan}}, \bibinfo {author} {\bibfnamefont {J.}~\bibnamefont {Zheng}},
  \bibinfo {author} {\bibfnamefont {Z.}~\bibnamefont {Chen}}, \bibinfo {author}
  {\bibfnamefont {W.}~\bibnamefont {Yang}}, \bibinfo {author} {\bibfnamefont
  {K.}~\bibnamefont {Amine}}, \ and\ \bibinfo {author} {\bibfnamefont
  {F.}~\bibnamefont {Pan}},\ }\href {\doibase 10.1021/acs.nanolett.5b04302}
  {\bibfield  {journal} {\bibinfo  {journal} {Nano Letters}\ }\textbf {\bibinfo
  {volume} {16}},\ \bibinfo {pages} {601} (\bibinfo {year} {2016})}\BibitemShut
  {NoStop}%
\bibitem [{\citenamefont {Chung}\ \emph {et~al.}(2002)\citenamefont {Chung},
  \citenamefont {Bloking},\ and\ \citenamefont {Chiang}}]{sung}%
  \BibitemOpen
  \bibfield  {author} {\bibinfo {author} {\bibfnamefont {S.-Y.}\ \bibnamefont
  {Chung}}, \bibinfo {author} {\bibfnamefont {J.~T.}\ \bibnamefont {Bloking}},
  \ and\ \bibinfo {author} {\bibfnamefont {Y.-M.}\ \bibnamefont {Chiang}},\
  }\href {\doibase 10.1038/nmat732} {\bibfield  {journal} {\bibinfo  {journal}
  {Nature Materials}\ }\textbf {\bibinfo {volume} {1}},\ \bibinfo {pages} {123}
  (\bibinfo {year} {2002})}\BibitemShut {NoStop}%
\bibitem [{\citenamefont {Ning}\ \emph {et~al.}(2014)\citenamefont {Ning},
  \citenamefont {Li}, \citenamefont {Xu},\ and\ \citenamefont {Ouyang}}]{ning}%
  \BibitemOpen
  \bibfield  {author} {\bibinfo {author} {\bibfnamefont {F.}~\bibnamefont
  {Ning}}, \bibinfo {author} {\bibfnamefont {S.}~\bibnamefont {Li}}, \bibinfo
  {author} {\bibfnamefont {B.}~\bibnamefont {Xu}}, \ and\ \bibinfo {author}
  {\bibfnamefont {C.}~\bibnamefont {Ouyang}},\ }\href {\doibase
  10.1016/j.ssi.2014.05.008} {\bibfield  {journal} {\bibinfo  {journal} {Solid
  State Ionics}\ }\textbf {\bibinfo {volume} {263}},\ \bibinfo {pages} {46}
  (\bibinfo {year} {2014})}\BibitemShut {NoStop}%
\bibitem [{\citenamefont {Bower}\ \emph {et~al.}(2011)\citenamefont {Bower},
  \citenamefont {Guduru},\ and\ \citenamefont {Sethuraman}}]{Bower}%
  \BibitemOpen
  \bibfield  {author} {\bibinfo {author} {\bibfnamefont {A.}~\bibnamefont
  {Bower}}, \bibinfo {author} {\bibfnamefont {P.}~\bibnamefont {Guduru}}, \
  and\ \bibinfo {author} {\bibfnamefont {V.}~\bibnamefont {Sethuraman}},\
  }\href {\doibase 10.1016/j.jmps.2011.01.003} {\bibfield  {journal} {\bibinfo
  {journal} {Journal of the Mechanics and Physics of Solids}\ }\textbf
  {\bibinfo {volume} {59}},\ \bibinfo {pages} {804} (\bibinfo {year}
  {2011})}\BibitemShut {NoStop}%
\bibitem [{\citenamefont {Garcia}\ \emph {et~al.}(2005)\citenamefont {Garcia},
  \citenamefont {Chiang}, \citenamefont {Carter}, \citenamefont {Limthongkul},\
  and\ \citenamefont {Bishop}}]{garcia}%
  \BibitemOpen
  \bibfield  {author} {\bibinfo {author} {\bibfnamefont {R.~E.}\ \bibnamefont
  {Garcia}}, \bibinfo {author} {\bibfnamefont {Y.-M.}\ \bibnamefont {Chiang}},
  \bibinfo {author} {\bibfnamefont {W.~C.}\ \bibnamefont {Carter}}, \bibinfo
  {author} {\bibfnamefont {P.}~\bibnamefont {Limthongkul}}, \ and\ \bibinfo
  {author} {\bibfnamefont {C.~M.}\ \bibnamefont {Bishop}},\ }\href {\doibase
  10.1149/1.1836132} {\bibfield  {journal} {\bibinfo  {journal} {Journal of The
  Electrochemical Society}\ }\textbf {\bibinfo {volume} {152}},\ \bibinfo
  {pages} {A255} (\bibinfo {year} {2005})}\BibitemShut {NoStop}%
\bibitem [{\citenamefont {Islam}\ and\ \citenamefont {Fisher}(2014)}]{Fisher2}%
  \BibitemOpen
  \bibfield  {author} {\bibinfo {author} {\bibfnamefont {M.~S.}\ \bibnamefont
  {Islam}}\ and\ \bibinfo {author} {\bibfnamefont {C.~A.~J.}\ \bibnamefont
  {Fisher}},\ }\href {\doibase 10.1039/c3cs60199d} {\bibfield  {journal}
  {\bibinfo  {journal} {Chem. Soc. Rev.}\ }\textbf {\bibinfo {volume} {43}},\
  \bibinfo {pages} {185} (\bibinfo {year} {2014})}\BibitemShut {NoStop}%
\bibitem [{\citenamefont {Okubo}\ \emph {et~al.}(2007)\citenamefont {Okubo},
  \citenamefont {Hosono}, \citenamefont {Kim}, \citenamefont {Enomoto},
  \citenamefont {Kojima}, \citenamefont {Kudo}, \citenamefont {Zhou},\ and\
  \citenamefont {Honma}}]{Okubo}%
  \BibitemOpen
  \bibfield  {author} {\bibinfo {author} {\bibfnamefont {M.}~\bibnamefont
  {Okubo}}, \bibinfo {author} {\bibfnamefont {E.}~\bibnamefont {Hosono}},
  \bibinfo {author} {\bibfnamefont {J.}~\bibnamefont {Kim}}, \bibinfo {author}
  {\bibfnamefont {M.}~\bibnamefont {Enomoto}}, \bibinfo {author} {\bibfnamefont
  {N.}~\bibnamefont {Kojima}}, \bibinfo {author} {\bibfnamefont
  {T.}~\bibnamefont {Kudo}}, \bibinfo {author} {\bibfnamefont {H.}~\bibnamefont
  {Zhou}}, \ and\ \bibinfo {author} {\bibfnamefont {I.}~\bibnamefont {Honma}},\
  }\href {\doibase 10.1021/ja0681927} {\bibfield  {journal} {\bibinfo
  {journal} {J. Am. Chem. Soc.}\ }\textbf {\bibinfo {volume} {129}},\ \bibinfo
  {pages} {7444} (\bibinfo {year} {2007})}\BibitemShut {NoStop}%
\bibitem [{\citenamefont {Choi}\ and\ \citenamefont {Pyun}(1997)}]{choi}%
  \BibitemOpen
  \bibfield  {author} {\bibinfo {author} {\bibfnamefont {Y.}~\bibnamefont
  {Choi}}\ and\ \bibinfo {author} {\bibfnamefont {S.}~\bibnamefont {Pyun}},\
  }\href {\doibase 10.1016/s0167-2738(97)00253-1} {\bibfield  {journal}
  {\bibinfo  {journal} {Solid State Ionics}\ }\textbf {\bibinfo {volume}
  {99}},\ \bibinfo {pages} {173} (\bibinfo {year} {1997})}\BibitemShut
  {NoStop}%
\bibitem [{\citenamefont {Wang}(1999)}]{wang2}%
  \BibitemOpen
  \bibfield  {author} {\bibinfo {author} {\bibfnamefont {H.}~\bibnamefont
  {Wang}},\ }\href {\doibase 10.1149/1.1391631} {\bibfield  {journal} {\bibinfo
   {journal} {Journal of The Electrochemical Society}\ }\textbf {\bibinfo
  {volume} {146}},\ \bibinfo {pages} {473} (\bibinfo {year}
  {1999})}\BibitemShut {NoStop}%
\bibitem [{\citenamefont {Takahashi}\ \emph {et~al.}(2007)\citenamefont
  {Takahashi}, \citenamefont {Kijima}, \citenamefont {Dokko}, \citenamefont
  {Nishizawa}, \citenamefont {Uchida},\ and\ \citenamefont
  {Akimoto}}]{Takahashi}%
  \BibitemOpen
  \bibfield  {author} {\bibinfo {author} {\bibfnamefont {Y.}~\bibnamefont
  {Takahashi}}, \bibinfo {author} {\bibfnamefont {N.}~\bibnamefont {Kijima}},
  \bibinfo {author} {\bibfnamefont {K.}~\bibnamefont {Dokko}}, \bibinfo
  {author} {\bibfnamefont {M.}~\bibnamefont {Nishizawa}}, \bibinfo {author}
  {\bibfnamefont {I.}~\bibnamefont {Uchida}}, \ and\ \bibinfo {author}
  {\bibfnamefont {J.}~\bibnamefont {Akimoto}},\ }\href {\doibase
  10.1016/j.jssc.2006.10.018} {\bibfield  {journal} {\bibinfo  {journal}
  {Journal of Solid State Chemistry}\ }\textbf {\bibinfo {volume} {180}},\
  \bibinfo {pages} {313} (\bibinfo {year} {2007})}\BibitemShut {NoStop}%
\bibitem [{\citenamefont {Diercks}\ \emph {et~al.}(2014)\citenamefont
  {Diercks}, \citenamefont {Musselman}, \citenamefont {Morgenstern},
  \citenamefont {Wilson}, \citenamefont {Kumar}, \citenamefont {Smith},
  \citenamefont {Kawase}, \citenamefont {Gorman}, \citenamefont {Eberhart},\
  and\ \citenamefont {Packard}}]{Diercks}%
  \BibitemOpen
  \bibfield  {author} {\bibinfo {author} {\bibfnamefont {D.~R.}\ \bibnamefont
  {Diercks}}, \bibinfo {author} {\bibfnamefont {M.}~\bibnamefont {Musselman}},
  \bibinfo {author} {\bibfnamefont {A.}~\bibnamefont {Morgenstern}}, \bibinfo
  {author} {\bibfnamefont {T.}~\bibnamefont {Wilson}}, \bibinfo {author}
  {\bibfnamefont {M.}~\bibnamefont {Kumar}}, \bibinfo {author} {\bibfnamefont
  {K.}~\bibnamefont {Smith}}, \bibinfo {author} {\bibfnamefont
  {M.}~\bibnamefont {Kawase}}, \bibinfo {author} {\bibfnamefont {B.~P.}\
  \bibnamefont {Gorman}}, \bibinfo {author} {\bibfnamefont {M.}~\bibnamefont
  {Eberhart}}, \ and\ \bibinfo {author} {\bibfnamefont {C.~E.}\ \bibnamefont
  {Packard}},\ }\href {\doibase 10.1149/2.0071411jes} {\bibfield  {journal}
  {\bibinfo  {journal} {Journal of the Electrochemical Society}\ }\textbf
  {\bibinfo {volume} {161}},\ \bibinfo {pages} {F3039} (\bibinfo {year}
  {2014})}\BibitemShut {NoStop}%
\bibitem [{\citenamefont {Kim}\ \emph {et~al.}(2004)\citenamefont {Kim},
  \citenamefont {Lee}, \citenamefont {Kim}, \citenamefont {Cho}, \citenamefont
  {Cho}, \citenamefont {Park}, \citenamefont {Oh},\ and\ \citenamefont
  {Yoon}}]{yong}%
  \BibitemOpen
  \bibfield  {author} {\bibinfo {author} {\bibfnamefont {Y.~J.}\ \bibnamefont
  {Kim}}, \bibinfo {author} {\bibfnamefont {E.-K.}\ \bibnamefont {Lee}},
  \bibinfo {author} {\bibfnamefont {H.}~\bibnamefont {Kim}}, \bibinfo {author}
  {\bibfnamefont {J.}~\bibnamefont {Cho}}, \bibinfo {author} {\bibfnamefont
  {Y.~W.}\ \bibnamefont {Cho}}, \bibinfo {author} {\bibfnamefont
  {B.}~\bibnamefont {Park}}, \bibinfo {author} {\bibfnamefont {S.~M.}\
  \bibnamefont {Oh}}, \ and\ \bibinfo {author} {\bibfnamefont {J.~K.}\
  \bibnamefont {Yoon}},\ }\href {\doibase 10.1149/1.1759611} {\bibfield
  {journal} {\bibinfo  {journal} {Journal of The Electrochemical Society}\
  }\textbf {\bibinfo {volume} {151}},\ \bibinfo {pages} {A1063} (\bibinfo
  {year} {2004})}\BibitemShut {NoStop}%
\bibitem [{\citenamefont {Mukhopadhyay}\ and\ \citenamefont
  {Sheldon}(2014)}]{sheldon}%
  \BibitemOpen
  \bibfield  {author} {\bibinfo {author} {\bibfnamefont {A.}~\bibnamefont
  {Mukhopadhyay}}\ and\ \bibinfo {author} {\bibfnamefont {B.~W.}\ \bibnamefont
  {Sheldon}},\ }\href {\doibase 10.1016/j.pmatsci.2014.02.001} {\bibfield
  {journal} {\bibinfo  {journal} {Progress in Materials Science}\ }\textbf
  {\bibinfo {volume} {63}},\ \bibinfo {pages} {58} (\bibinfo {year}
  {2014})}\BibitemShut {NoStop}%
\bibitem [{\citenamefont {Balke}\ \emph {et~al.}(2010)\citenamefont {Balke},
  \citenamefont {Jesse}, \citenamefont {Morozovska}, \citenamefont {Eliseev},
  \citenamefont {Chung}, \citenamefont {Kim}, \citenamefont {Adamczyk},
  \citenamefont {Garc{\'{\i}}a}, \citenamefont {Dudney},\ and\ \citenamefont
  {Kalinin}}]{Balke}%
  \BibitemOpen
  \bibfield  {author} {\bibinfo {author} {\bibfnamefont {N.}~\bibnamefont
  {Balke}}, \bibinfo {author} {\bibfnamefont {S.}~\bibnamefont {Jesse}},
  \bibinfo {author} {\bibfnamefont {A.~N.}\ \bibnamefont {Morozovska}},
  \bibinfo {author} {\bibfnamefont {E.}~\bibnamefont {Eliseev}}, \bibinfo
  {author} {\bibfnamefont {D.~W.}\ \bibnamefont {Chung}}, \bibinfo {author}
  {\bibfnamefont {Y.}~\bibnamefont {Kim}}, \bibinfo {author} {\bibfnamefont
  {L.}~\bibnamefont {Adamczyk}}, \bibinfo {author} {\bibfnamefont {R.~E.}\
  \bibnamefont {Garc{\'{\i}}a}}, \bibinfo {author} {\bibfnamefont
  {N.}~\bibnamefont {Dudney}}, \ and\ \bibinfo {author} {\bibfnamefont {S.~V.}\
  \bibnamefont {Kalinin}},\ }\href {\doibase 10.1038/nnano.2010.174} {\bibfield
   {journal} {\bibinfo  {journal} {Nature Nanotech}\ }\textbf {\bibinfo
  {volume} {5}},\ \bibinfo {pages} {749} (\bibinfo {year} {2010})}\BibitemShut
  {NoStop}%
\bibitem [{\citenamefont {Xiong}\ \emph {et~al.}(2012)\citenamefont {Xiong},
  \citenamefont {Yan}, \citenamefont {Chen}, \citenamefont {Xu}, \citenamefont
  {Le},\ and\ \citenamefont {Ouyang}}]{ouyang2}%
  \BibitemOpen
  \bibfield  {author} {\bibinfo {author} {\bibfnamefont {F.}~\bibnamefont
  {Xiong}}, \bibinfo {author} {\bibfnamefont {H.~J.}\ \bibnamefont {Yan}},
  \bibinfo {author} {\bibfnamefont {Y.}~\bibnamefont {Chen}}, \bibinfo {author}
  {\bibfnamefont {B.}~\bibnamefont {Xu}}, \bibinfo {author} {\bibfnamefont
  {J.~X.}\ \bibnamefont {Le}}, \ and\ \bibinfo {author} {\bibfnamefont {C.~Y.}\
  \bibnamefont {Ouyang}},\ }\href@noop {} {\bibfield  {journal} {\bibinfo
  {journal} {Int. J. Electrochem. Sci.}\ }\textbf {\bibinfo {volume} {7}},\
  \bibinfo {pages} {9390 } (\bibinfo {year} {2012})}\BibitemShut {NoStop}%
\bibitem [{\citenamefont {Li}\ \emph {et~al.}(2009)\citenamefont {Li},
  \citenamefont {Cheng},\ and\ \citenamefont {Zhang}}]{Li}%
  \BibitemOpen
  \bibfield  {author} {\bibinfo {author} {\bibfnamefont {Y.}~\bibnamefont
  {Li}}, \bibinfo {author} {\bibfnamefont {X.}~\bibnamefont {Cheng}}, \ and\
  \bibinfo {author} {\bibfnamefont {Y.}~\bibnamefont {Zhang}},\ }in\ \href
  {\doibase 10.1149/1.3248343} {\emph {\bibinfo {booktitle} {{ECS}
  Transactions}}}\ (\bibinfo  {publisher} {The Electrochemical Society},\
  \bibinfo {year} {2009})\BibitemShut {NoStop}%
\bibitem [{\citenamefont {Renganathan}\ \emph {et~al.}(2010)\citenamefont
  {Renganathan}, \citenamefont {Sikha}, \citenamefont {Santhanagopalan},\ and\
  \citenamefont {White}}]{Renganathan}%
  \BibitemOpen
  \bibfield  {author} {\bibinfo {author} {\bibfnamefont {S.}~\bibnamefont
  {Renganathan}}, \bibinfo {author} {\bibfnamefont {G.}~\bibnamefont {Sikha}},
  \bibinfo {author} {\bibfnamefont {S.}~\bibnamefont {Santhanagopalan}}, \ and\
  \bibinfo {author} {\bibfnamefont {R.~E.}\ \bibnamefont {White}},\ }\href
  {\doibase 10.1149/1.3261809} {\bibfield  {journal} {\bibinfo  {journal}
  {Journal of The Electrochemical Society}\ }\textbf {\bibinfo {volume}
  {157}},\ \bibinfo {pages} {A155} (\bibinfo {year} {2010})}\BibitemShut
  {NoStop}%
\bibitem [{\citenamefont {Zhao}\ \emph {et~al.}(2010)\citenamefont {Zhao},
  \citenamefont {Pharr}, \citenamefont {Vlassak},\ and\ \citenamefont
  {Suo}}]{zhao}%
  \BibitemOpen
  \bibfield  {author} {\bibinfo {author} {\bibfnamefont {K.}~\bibnamefont
  {Zhao}}, \bibinfo {author} {\bibfnamefont {M.}~\bibnamefont {Pharr}},
  \bibinfo {author} {\bibfnamefont {J.~J.}\ \bibnamefont {Vlassak}}, \ and\
  \bibinfo {author} {\bibfnamefont {Z.}~\bibnamefont {Suo}},\ }\href {\doibase
  10.1063/1.3492617} {\bibfield  {journal} {\bibinfo  {journal} {J. Appl.
  Phys.}\ }\textbf {\bibinfo {volume} {108}},\ \bibinfo {pages} {073517}
  (\bibinfo {year} {2010})}\BibitemShut {NoStop}%
\bibitem [{\citenamefont {Leibfried}\ and\ \citenamefont
  {Breuer}(1978)}]{Leibfried}%
  \BibitemOpen
  \bibfield  {author} {\bibinfo {author} {\bibfnamefont {G.}~\bibnamefont
  {Leibfried}}\ and\ \bibinfo {author} {\bibfnamefont {N.}~\bibnamefont
  {Breuer}},\ }\href@noop {} {\emph {\bibinfo {title} {Point defects in metals
  I}}}\ (\bibinfo  {publisher} {Springer Verlag},\ \bibinfo {year}
  {1978})\BibitemShut {NoStop}%
\bibitem [{\citenamefont {Varvenne}\ \emph {et~al.}(2013)\citenamefont
  {Varvenne}, \citenamefont {Bruneval}, \citenamefont {Marinica},\ and\
  \citenamefont {Clouet}}]{Varvenne2013}%
  \BibitemOpen
  \bibfield  {author} {\bibinfo {author} {\bibfnamefont {C.}~\bibnamefont
  {Varvenne}}, \bibinfo {author} {\bibfnamefont {F.}~\bibnamefont {Bruneval}},
  \bibinfo {author} {\bibfnamefont {M.-C.}\ \bibnamefont {Marinica}}, \ and\
  \bibinfo {author} {\bibfnamefont {E.}~\bibnamefont {Clouet}},\ }\href
  {\doibase 10.1103/physrevb.88.134102} {\bibfield  {journal} {\bibinfo
  {journal} {Physical Review B}\ }\textbf {\bibinfo {volume} {88}} (\bibinfo
  {year} {2013}),\ 10.1103/physrevb.88.134102}\BibitemShut {NoStop}%
\bibitem [{\citenamefont {Leslie}\ and\ \citenamefont
  {Gillan}(1985)}]{gillan3}%
  \BibitemOpen
  \bibfield  {author} {\bibinfo {author} {\bibfnamefont {M.}~\bibnamefont
  {Leslie}}\ and\ \bibinfo {author} {\bibfnamefont {N.~J.}\ \bibnamefont
  {Gillan}},\ }\href {\doibase 10.1088/0022-3719/18/5/005} {\bibfield
  {journal} {\bibinfo  {journal} {J. Phys. C: Solid State Phys.}\ }\textbf
  {\bibinfo {volume} {18}},\ \bibinfo {pages} {973} (\bibinfo {year}
  {1985})}\BibitemShut {NoStop}%
\bibitem [{\citenamefont {Zhang}\ and\ \citenamefont
  {Northrup}(1991)}]{Zhang1991}%
  \BibitemOpen
  \bibfield  {author} {\bibinfo {author} {\bibfnamefont {S.}~\bibnamefont
  {Zhang}}\ and\ \bibinfo {author} {\bibfnamefont {J.}~\bibnamefont
  {Northrup}},\ }\href {\doibase 10.1103/physrevlett.67.2339} {\bibfield
  {journal} {\bibinfo  {journal} {Physical Review Letters}\ }\textbf {\bibinfo
  {volume} {67}},\ \bibinfo {pages} {2339} (\bibinfo {year}
  {1991})}\BibitemShut {NoStop}%
\bibitem [{\citenamefont {der Ven}\ and\ \citenamefont
  {Ceder}(1999)}]{VanderVen}%
  \BibitemOpen
  \bibfield  {author} {\bibinfo {author} {\bibfnamefont {A.~V.}\ \bibnamefont
  {der Ven}}\ and\ \bibinfo {author} {\bibfnamefont {G.}~\bibnamefont
  {Ceder}},\ }\href {\doibase 10.1149/1.1391130} {\bibfield  {journal}
  {\bibinfo  {journal} {Electrochemical and Solid-State Letters}\ }\textbf
  {\bibinfo {volume} {3}},\ \bibinfo {pages} {301} (\bibinfo {year}
  {1999})}\BibitemShut {NoStop}%
\bibitem [{\citenamefont {http://www.cs.sandia.gov/
  paschul/Quest/}(2015)}]{seq}%
  \BibitemOpen
  \bibfield  {author} {\bibinfo {author} {\bibnamefont
  {http://www.cs.sandia.gov/ paschul/Quest/}},\ }\href@noop {} {} (\bibinfo
  {year} {2015})\BibitemShut {NoStop}%
\bibitem [{\citenamefont {Perdew}\ \emph {et~al.}(1996)\citenamefont {Perdew},
  \citenamefont {Burke},\ and\ \citenamefont {Ernzerhof}}]{PBE}%
  \BibitemOpen
  \bibfield  {author} {\bibinfo {author} {\bibfnamefont {J.~P.}\ \bibnamefont
  {Perdew}}, \bibinfo {author} {\bibfnamefont {K.}~\bibnamefont {Burke}}, \
  and\ \bibinfo {author} {\bibfnamefont {M.}~\bibnamefont {Ernzerhof}},\ }\href
  {\doibase 10.1103/physrevlett.77.3865} {\bibfield  {journal} {\bibinfo
  {journal} {Phys. Rev. Lett.}\ }\textbf {\bibinfo {volume} {77}},\ \bibinfo
  {pages} {3865} (\bibinfo {year} {1996})}\BibitemShut {NoStop}%
\bibitem [{\citenamefont {J{\'o}nsson}\ \emph {et~al.}(1998)\citenamefont
  {J{\'o}nsson}, \citenamefont {Mills},\ and\ \citenamefont {Jacobsen}}]{NEB}%
  \BibitemOpen
  \bibfield  {author} {\bibinfo {author} {\bibfnamefont {H.}~\bibnamefont
  {J{\'o}nsson}}, \bibinfo {author} {\bibfnamefont {G.}~\bibnamefont {Mills}},
  \ and\ \bibinfo {author} {\bibfnamefont {K.~W.}\ \bibnamefont {Jacobsen}},\
  }\href@noop {} {\  (\bibinfo {year} {1998})}\BibitemShut {NoStop}%
\bibitem [{\citenamefont {Kresse}\ and\ \citenamefont
  {Furthm{\"u}ller}(1996)}]{vasp}%
  \BibitemOpen
  \bibfield  {author} {\bibinfo {author} {\bibfnamefont {G.}~\bibnamefont
  {Kresse}}\ and\ \bibinfo {author} {\bibfnamefont {J.}~\bibnamefont
  {Furthm{\"u}ller}},\ }\href@noop {} {\bibfield  {journal} {\bibinfo
  {journal} {Phys. Rev. B}\ }\textbf {\bibinfo {volume} {54}},\ \bibinfo
  {pages} {169} (\bibinfo {year} {1996})}\BibitemShut {NoStop}%
\bibitem [{\citenamefont {Zhang}\ \emph {et~al.}(2013)\citenamefont {Zhang},
  \citenamefont {Li}, \citenamefont {Tao},\ and\ \citenamefont {Chen}}]{chen}%
  \BibitemOpen
  \bibfield  {author} {\bibinfo {author} {\bibfnamefont {T.}~\bibnamefont
  {Zhang}}, \bibinfo {author} {\bibfnamefont {D.}~\bibnamefont {Li}}, \bibinfo
  {author} {\bibfnamefont {Z.}~\bibnamefont {Tao}}, \ and\ \bibinfo {author}
  {\bibfnamefont {J.}~\bibnamefont {Chen}},\ }\href {\doibase
  10.1016/j.pnsc.2013.04.005} {\bibfield  {journal} {\bibinfo  {journal}
  {Progress in Natural Science: Materials International}\ }\textbf {\bibinfo
  {volume} {23}},\ \bibinfo {pages} {256} (\bibinfo {year} {2013})}\BibitemShut
  {NoStop}%
\bibitem [{\citenamefont {Zhu}\ \emph {et~al.}(2014)\citenamefont {Zhu},
  \citenamefont {Liu},\ and\ \citenamefont {Scarpulla}}]{Zhu2014}%
  \BibitemOpen
  \bibfield  {author} {\bibinfo {author} {\bibfnamefont {J.}~\bibnamefont
  {Zhu}}, \bibinfo {author} {\bibfnamefont {F.}~\bibnamefont {Liu}}, \ and\
  \bibinfo {author} {\bibfnamefont {M.~A.}\ \bibnamefont {Scarpulla}},\ }\href
  {\doibase 10.1063/1.4863076} {\bibfield  {journal} {\bibinfo  {journal}
  {{APL} Materials}\ }\textbf {\bibinfo {volume} {2}},\ \bibinfo {pages}
  {012110} (\bibinfo {year} {2014})}\BibitemShut {NoStop}%
\bibitem [{\citenamefont {Aschauer}\ \emph {et~al.}(2015)\citenamefont
  {Aschauer}, \citenamefont {Vonr\"{u}ti},\ and\ \citenamefont
  {Spaldin}}]{Aschauer2015}%
  \BibitemOpen
  \bibfield  {author} {\bibinfo {author} {\bibfnamefont {U.}~\bibnamefont
  {Aschauer}}, \bibinfo {author} {\bibfnamefont {N.}~\bibnamefont
  {Vonr\"{u}ti}}, \ and\ \bibinfo {author} {\bibfnamefont {N.~A.}\ \bibnamefont
  {Spaldin}},\ }\href {\doibase 10.1103/physrevb.92.054103} {\bibfield
  {journal} {\bibinfo  {journal} {Physical Review B}\ }\textbf {\bibinfo
  {volume} {92}} (\bibinfo {year} {2015}),\
  10.1103/physrevb.92.054103}\BibitemShut {NoStop}%
\end{thebibliography}%

\end{document}